\documentclass[prd,twocolumn,nofootinbib,superscriptaddress]{revtex4-2}

\usepackage{preamble}

\begin{document}

\title{Mathematical framework for simulation of non-perturbative cavity optomechanics with plate-like dielectrics
}

\author{Andrew Laeuger\,\orcidlink{0000-0002-8212-6496}}
\email{alaeuger@caltech.edu}
\affiliation{Department of Physics and Astronomy, Caltech, Pasadena, California 91125, USA}

\author{Nancy Aggarwal}
\email{nqaggarwal@ucdavis.edu}
\affiliation{Department of Physics and Astronomy, University of California, Davis, California 95616, USA}

\date{\today}

\begin{abstract}
Cavity optomechanical models typically treat a sensor either as a perturbative polarizable object in a pre-existing optical field, or solve the field at a fixed equilibrium and linearize in a fixed optical-mode basis. Neither self-consistently determines the spatially resolved cavity field, the optical force, the mechanical equilibrium, and the back-action when a finite-thickness dielectric plate intercepts the entire optical mode and restructures the intracavity field.
In this companion paper to \cite{DielectricLetter}, we develop that treatment by extending the vacuum two-photon formalism to finite-thickness planar dielectrics, incorporating their Fresnel reflection, internal propagation, optical forces, and mechanical back-action directly into the field equations.
We derive the carrier and sideband transfer functions needed to describe the coupled optical-mechanical system, including optomechanical cooling and the response to external perturbations. 
The result is a systematic starting point for calculating detector response and noise budgets for cavity experiments with intracavity dielectrics, including gravitational wave detectors deploying levitated sensors. 

\end{abstract}

\maketitle

Recent proposals for gravitational wave detectors at high frequencies using levitated sensors \cite{LSD2012,LSD2020,AndrewsGRNote} employ particles with wide, plate-like geometries,\footnote{More recently, dielectric intracavity membranes have also been proposed as an avenue for high frequency GW detection \cite{Rousso:2026duo}.} and need quantitative predictions for the whole cavity-sensor system: the optical field, the force it exerts on the sensor, the equilibrium it sets, and the signal and noise transfer functions that follow. Recent experimental work has demonstrated
the successful levitation of a high aspect-ratio single ma-
terial dielectric in a dual-beam optical trap \cite{LSD2022}, and efforts
are ongoing to levitate single and multilayer dielectrics
of even larger cross sections \cite{DiskTrappingPaper}.

The available tools come from the broad body of work on optical resonators coupled to mechanical motion. The strong dependence of cavity fields on mechanical position enables optical readout of minute displacements while radiation pressure provides a means of carefully manipulating the mechanical state. Experiments built on these capabilities have probed radiation-pressure backaction, laser cooling of mechanical motion, quantum behavior of macroscopic objects, and gravitational-wave detection, and have reached increadingly sensitive measurements of minute forces, torques, and accelerations \cite{CavityOptomechanics, Kippenburg2007, Li2021,cripe2019measurement, Metcalfe2014,Cripe:2017cvb, Marquardt2009,aggarwal2020room, Gavartin2012, Ranjit:2016dlq, Liang2023, Guha2020, Komori2020, Zhou2021, Ahn:2019cbu, Ju:2023egl, Dana2026}. Within this body of work, intracavity dielectrics enter the model in two main ways.

In levitated optomechanics, nano-scale dielectric particles can be levitated in regions of large optical intensity via the gradient force \cite{ChangSpheres, ChangThinDiscs, Millen2020}. Despite the small size of the test mass, their decoupling from the common environmental noise sources has led them to be exquisite force sensors, including searches for non-Newtonian gravitational interactions and Casimir effects \cite{Ranjit:2016dlq, Montoya:2021dqz, Geraci:2010ft, Monteiro:2020qiz, Hebestreit:2017xur, Kawasaki:2020oyl}. These systems are often modeled by treating the particle as a perturbative polarizable object whose position-dependent interaction with a pre-existing optical field determines its optomechanical coupling \cite{ChangSpheres, Millen2020, Maurer:2021zgx, Maurer:2022nvj, Reigue:2022thv}. This treatment does not allow the levitated object to reshape the field that traps it.

Dielectric membranes on the other hand, mechanically mounted along the cavity axis, capture the entire optical field profile and are typically highly reflective \cite{jayich2008dispersive, Thompson:2007hno}.  They are often handled via scattering or transfer matrix methods which explicitly incorporate the dielectric without perturbative expansions  \cite{jayich2008dispersive,xuereb2010optomechanical,xuereb2012dynamical}. In the high-reflectivity limit, such treatments reveal that the membrane can effectively form two weakly-coupled subcavities, producing normal-mode splitting and avoided crossings between symmetric and antisymmetric optical modes \cite{ Dumont:2019ziy, DumontThesis, stambaugh2015membrane}. Related multi-mode optomechanical frameworks have been developed to analyze optical spring effects and other closed-loop dynamics, along with quantum noise and measurement sensitivity \cite{xu2014squeezing, Rabl:2011nyt}. In these treatments, however, the optical field is typically first determined for a given equilibrium configuration and represented in a fixed optical-mode basis, after which mechanical motion is incorporated through position-dependent changes in the mode frequencies and couplings. This approach captures the resulting optomechanical dynamics through linearization about the equilibrium, but does not require that the reorganization of the spatial optical field to be computed self-consistently as the dielectric moves.

A distinct regime arises when a freely levitated dielectric captures the full optical beam cross section while remaining only moderately reflective. The dielectric can then substantially restructure the intracavity field, no longer simply behaving as a small perturbatiion on the field, while its equilibrium position is itself determined by the resulting optical forces. Neither of the two treatments above is built for this regime.

Concretely, a model that can support detector design needs four ingredients. \textit{(i)} The field along the cavity axis must be solved for any sensor position, without assuming a fixed set of optical modes. \textit{(ii)} The DC and AC optical forces must be computed from those same fields, so that the equilibrium and the trap stiffness follow from them. \textit{(iii)} The sideband response must be tracked in both quadratures, including the back-action of the trap and probe beams on the sensor and on each other. \textit{(iv)} Perturbations must be accepted at every port of the experiment, including laser fluctuations, mirror displacements, external forces, and a gravitational wave, so that signal and noise transfer functions can be assembled into a noise budget. The treatments above either take the field as given or determine it for a fixed equilibrium; the framework developed here supplies all four.

The formalism we build on is the two-photon formalism \cite{CavesAndSchumaker1,CavesAndSchumaker2,CorbittFramework}, in which an experiment is a chain of elements acting on carrier and sideband quadrature vectors: free space, moving mirrors, and beamsplitters. A moving dielectric of finite thickness is the missing element. Its two faces and its internal propagation must be treated together, and its motion changes the very fields that exert force on it. Supplying this element, along with the optical force and a mechanical susceptibility, is what allows the formalism to describe interferometric detectors integrated with levitated sensors.

The need for a self-consistent treatment is not simply a formality. 
In an accompanying letter \cite{DielectricLetter}, we show that a thin, moderately reflective plate partitions the cavity into two highly-coupled subcavities, producing new optical resonances and trapping sites with locations and stiffnesses far from the small-particle limit. We further find that conventional perturbative expressions for the cavity frequency shift and optomechanical coupling can depart significantly from the exact solution for sensor thicknesses as small as $\sim\lambda/(10n)$, while cavity locking can decrease the optomechanical coupling and increase the trap stiffness by orders of magnitude. Finally, the subcavity structure enables appreciable optical cooling even at zero detuning, a sharp contrast from the usual single-mode perturbative picture. These results establish a distinct regime of cavity optomechanics in which the sensor itself helps define the intracavity field, highlighting the importance of the self-consistent framework at the center of this paper which properly accounts for the coupled optical fields, radiation-pressure forces, and mechanical response.

\begin{figure}
    \centering
    \includegraphics[width=0.95\linewidth]{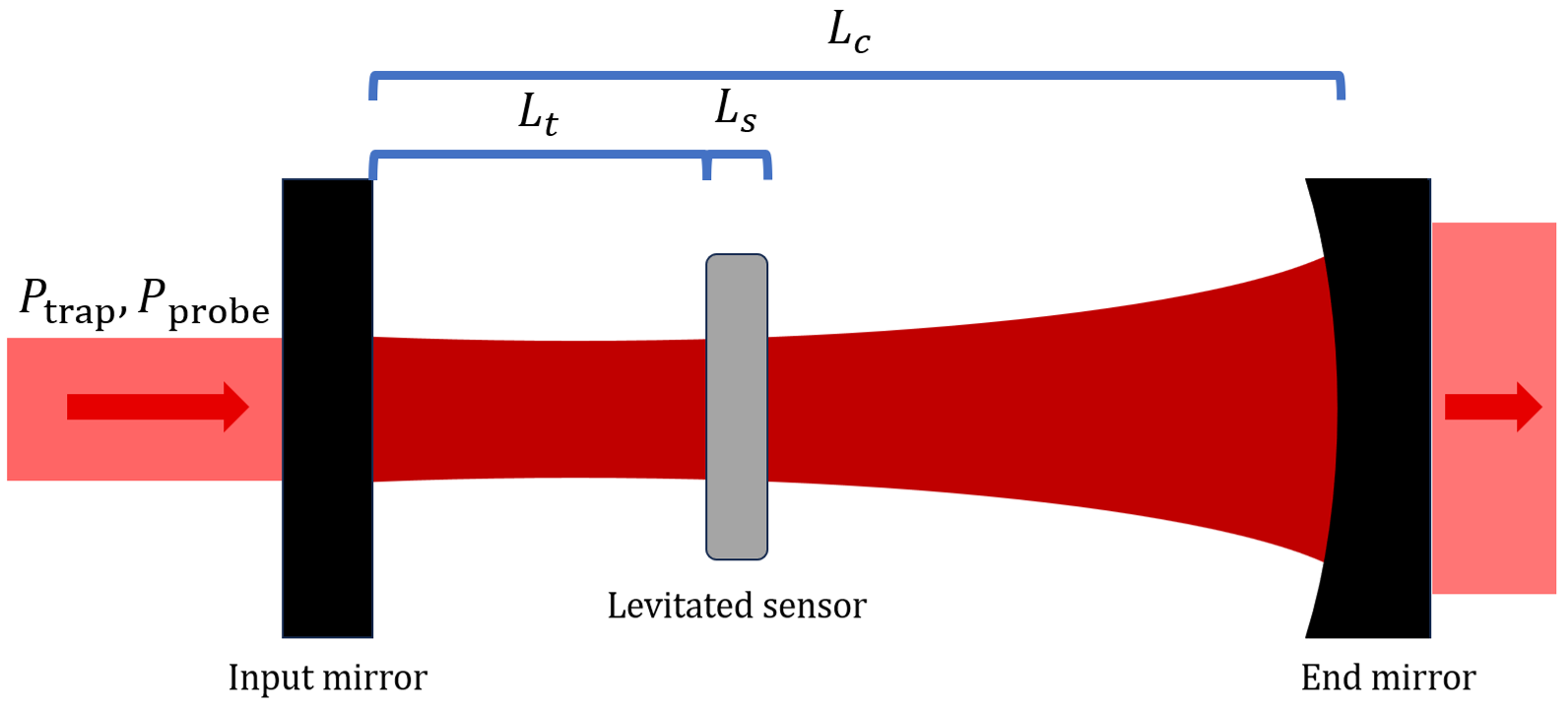}
    \caption{Basic diagram of a plate-like levitated sensor trapped within a cavity. A spatially similar trap and probe beam, with powers $P_\text{trap}$ and $P_\text{probe}$, respectively, enter the cavity of length $L_c$ from the left. The planar sensor, with thickness $L_s$, is trapped a distance $L_t$ away from the input mirror. Darker red shades indicate greater laser intensities. The sensor's cross section captures nearly the entire beam profile for both the trap and probe lasers at the trap location. The probe beam's primary use is readout of the sensor position, but it also cools the sensor's motion along the cavity axis.}
    \label{fig: experiment diagram cavity only}
\end{figure}

In Fig. \ref{fig: experiment diagram cavity only}, we present a schematic diagram of the simplest form of such an experiment. In this system, two coaxial lasers --- a strong trap beam and a weak probe, with respective powers $P_\text{trap}$ and $P_\text{probe}$ --- enter an optical cavity of length\footnote{We use ``$\approx$" because we will encode sub-wavelength changes in macroscopic lengths via a separate phase parameter, see Fig. \ref{fig:fields diagram complete}.} $\approx L_c$. The trap beam provides the optical forces which levitate a planar dielectric sensor of thickness $L_s$ a distance $\approx L_t$ from the cavity input mirror. The position of the sensor along the cavity axis is encoded in the phase of the probe light leaving the cavity back through the input mirror, though the probe beam also serves to cool the sensor's thermal motion along the cavity axis. Despite its simplicity, this model will allow for the presentation of all the key features of our framework --- more complicated experimental designs can be constructed with the tools outlined here.

As indicated by the diagram, the input mirror is flat and the end mirror is curved so that the waists of the trap and probe Gaussian beam profiles lie at the input mirror. This configuration places the optimal trap sites near the input mirror, where the response from an incident GW is maximized \cite{AndrewsGRNote}. However, the formalism we develop in this paper is completely general in the sensor position, only assuming that the dielectric cross section captures most of both the trap and probe beam powers where it is located. This assumption implies that scattering from the levitated sensor will occur primarily along the cavity axis. Therefore, when we study the propagation of optical fields, we employ the assumption that these fields are well-described by plane waves propagating between surfaces of infinite areal extent, all of which are aligned with the cavity axis. It is worth noting sensors with $L_s$ equal to an integer multiple of half the reduced wavelength are completely transparent under the plane wave approximation. Thus, our model cannot capture the residual trapping physics at these special thicknesses, and a treatment including the transverse beam profile is required. Throughout this work, we also assume that all optics are lossless, though extending our results to the nonzero-loss case is straightforward.

The remainder of this work is organized around the formalism itself.
In Sec. \ref{sec: cavity equations and optomechanical loops}, we extend the field propagation relations and optical forces of the two-photon formalism to a planar dielectric sensor and outline the computation of self-consistent response functions, cooling via optical back-action, and sensitivity estimates. The form of the optical force on the dielectric slab features unintuitive subtleties, which are derived and addressed in detail in a companion paper \cite{ForceLetter}. In Sec. \ref{sec: detector response}, we compute numerical experimental response functions and study their structure in detail. We conclude in Sec. \ref{sec: conclusion}.

\section{Two-photon formalism with planar levitated dielectrics}
\label{sec: cavity equations and optomechanical loops}

In this section, we outline the formalism we employ to describe the propagation of light fields within this experiment, which generally follows that of \cite{CorbittFramework} but is extended to incorporate the physics needed to properly model levitated dielectric sensors. We then show how the same formalism incorporates the optical forces, mechanical response, and optomechanical back-action needed to construct self-consistent transfer functions. These ingredients provide the basis for the detector-response calculations developed below.

In Tab. \ref{tab:parameters}, we provide a reference for variable names used throughout this section and in the remainder of this work. Non-empty entries in the ``Value" column indicate values used in the numerical examples.

\begin{table}
\centering
\textbf{Reference Table}\\
\vspace*{0.15cm}
\begin{tabular}{lcc}
\hline
\hline
Parameter & Symbol & Value \\ 
\hline
Trap laser wavenumber & $k$ & $2\pi/(1550$ nm) \\
Laser frequency & $\omega_0$ & $c\times k$ \\
Trap laser power & $P_\text{trap}$ & 1 W\\
Probe laser power & $P_\text{probe}$ & 10 mW\\
Cavity length & $L_c$ & 1 m \\
Mirror amp. reflectivity & $\rho$ & $\sqrt{0.96}$,
\\
Mirror amp. transmissivity & $\tau$ & $\equiv\sqrt{1-\rho^2}$ \\
Trap distance & $L_t$ & 1 mm \\
Cavity linewidth & $\kappa$ & \\
Probe detuning freq. & $\Delta$ & \\
Sensor thickness & $L_s$ &  \\
Sensor index of refraction & $n$ & 1.45 (silica) \\
Probe wavevector in sensor & $\bar{k}$ & $n(k+\Delta/c)$  \\
Sensor front face phase & $\phi_f$ & \\
Sensor radius & $R$ & 75 $\mu$m \\
Sensor mass & $m$ & 50 ng$\times L_s/(\lambda/n)$\\
- & - & (silica: 2.65 g/$\text{cm}^3$) \\
End mirror face phase & $\phi_e$ & \\
Beam waist radius & $w_0$ & 37.5$\mu$m \\
Sideband (GW) frequency & $\Omega$ & \\
Trap resonant frequency & $\Omega_\text{trap}$ & \\
Gas damping rate & $\Gamma_\text{gas}$ & $32PR^2/\bar vm$ \cite{LSD2020}\\
Temperature & $T$ & 300 K \\
Pressure & $P$ & $10^{-9}$ torr \\
\hline
\end{tabular}
\caption{Variables and parameters used throughout this paper. Completed ``Value" entries indicate that the parameter is fixed for all numerical results where not explicitly stated.}
\label{tab:parameters}
\end{table}

\subsection{Preliminaries}
\label{sec: quadrature formalism preliminaries}
We assume an electric field $\vec{\mathbb{E}}$ which is polarized purely in the $z-$direction:
\begin{equation}
    \vec{\mathbb{E}}(x,y,z,t)=f(y,z)\mathcal{E}(x,t)\hat z,
\end{equation}
where $f(y,z)$ is the cross-sectional distribution of the field amplitude. We assume a Gaussian cross-section; this choice appears only in the scalar prefactor for force expressions. Because the reflecting surfaces in the experiment all have radii larger than the local beam waist, we can safely neglect $f(y,z)$ until the calculation of forces.

At any point along the cavity axis, we first decompose $\mathcal{E}$ into a right- ($\mathcal{E}_+$) and left-propagating ($\mathcal{E}_-$) component, writing $\mathcal{E}_\pm(x,t)$ 
\begin{equation}
    \mathcal{E}_\pm(x,t)=\mathcal{E}_{\pm,1}(x,t)\cos(\omega_0 t)+\mathcal{E}_{\pm,2}(x,t)\sin(\omega_0 t),
\end{equation}
where $\omega_0$ is the carrier frequency of the field and $\mathcal{E}_{\pm,1/2}$ are the amplitude and phase quadrature components of the field. The time-evolution of the quadrature components is assumed to be slow relative to $\omega_0$. It is useful to define
\begin{equation}
    \vec{\mathcal{E}}_\pm(x,t)=\begin{pmatrix}
        \mathcal{E}_{\pm,1}(x,t) \\ \mathcal{E}_{\pm,2}(x,t)
    \end{pmatrix},
\end{equation}
which will reduce the complexity of future expressions.

This quadrature decomposition can be extended to the carrier-sideband picture, as originally outlined by Caves and Schumaker \cite{CavesAndSchumaker1,CavesAndSchumaker2} and further adapted to optical cavities in \cite{CorbittFramework}. The two quadratures are broken into a constant and fluctuating component, i.e.,
\begin{equation}
    \mathcal{E}_{\pm,1/2}(x,t)=E_{\pm,1/2}(x)+e_{\pm,1/2}(x,t).
\end{equation}
We then write $e_{\pm,1/2}(x,\Omega)$ as the Fourier transform of $e_{\pm,1/2}(x,t)$, using the convention
\begin{equation}
    e_{\pm,1/2}(x,\Omega)\equiv \int_{-\infty}^\infty e_{\pm,1/2}(x,t)\exp(i\Omega t)dt.
\end{equation}
The constant part gives the carrier field amplitude, while the fluctuating component is the sideband field. Much like the quadrature vector $\vec{\mathcal{E}}_\pm$ defined earlier, we write
\begin{equation}
    \vec E_\pm(x)=\begin{pmatrix}
        E_{\pm,1}(x) \\ E_{\pm,2}(x)
    \end{pmatrix},\quad 
    \vec e_\pm(x,\Omega)=\begin{pmatrix}
        e_{\pm,1}(x,\Omega) \\ e_{\pm,2}(x,\Omega)
    \end{pmatrix}.
\end{equation}
Throughout this work, we often remove the dependence on the cavity axis coordinate, along with the $\pm$ subscript, instead conveying this information with unique Latin alphabet quadrature vectors. Thus, when a vector label is attached to standard Latin alphabet variables (e.g., $\vec C$, $\vec h$, $\vec e^\text{ tr}$), they refer to carrier or sideband vectors evaluated at a predetermined point along the cavity axis and with a predetermined propagation direction, both of which can be found by referring to the relevant figures.

\subsection{Formalism for electric field propagation}
We now turn to describing how these quadrature vectors evolve across simple elements of an optical cavity: free space, a moving mirror, and a moving planar dielectric of finite thickness. For more complex interferometer geometries, a beamsplitter element may be useful as well; analogous expressions for such a component to those presented below appear in \cite{CorbittFramework}. Thorough derivations of the results for free space and a moving mirror can be found in \cite{AggarwalThesis}; here, we just quote them in their simplest form.

\begin{figure}
    \centering
    \begin{minipage}{0.48\columnwidth}
        \centering
        \includegraphics[width=\textwidth]{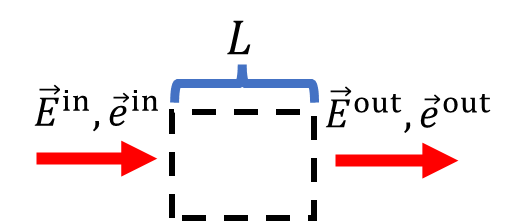}
    \end{minipage}
    \hfill
    \begin{minipage}{0.48\columnwidth}
        \centering
        \includegraphics[width=\textwidth]{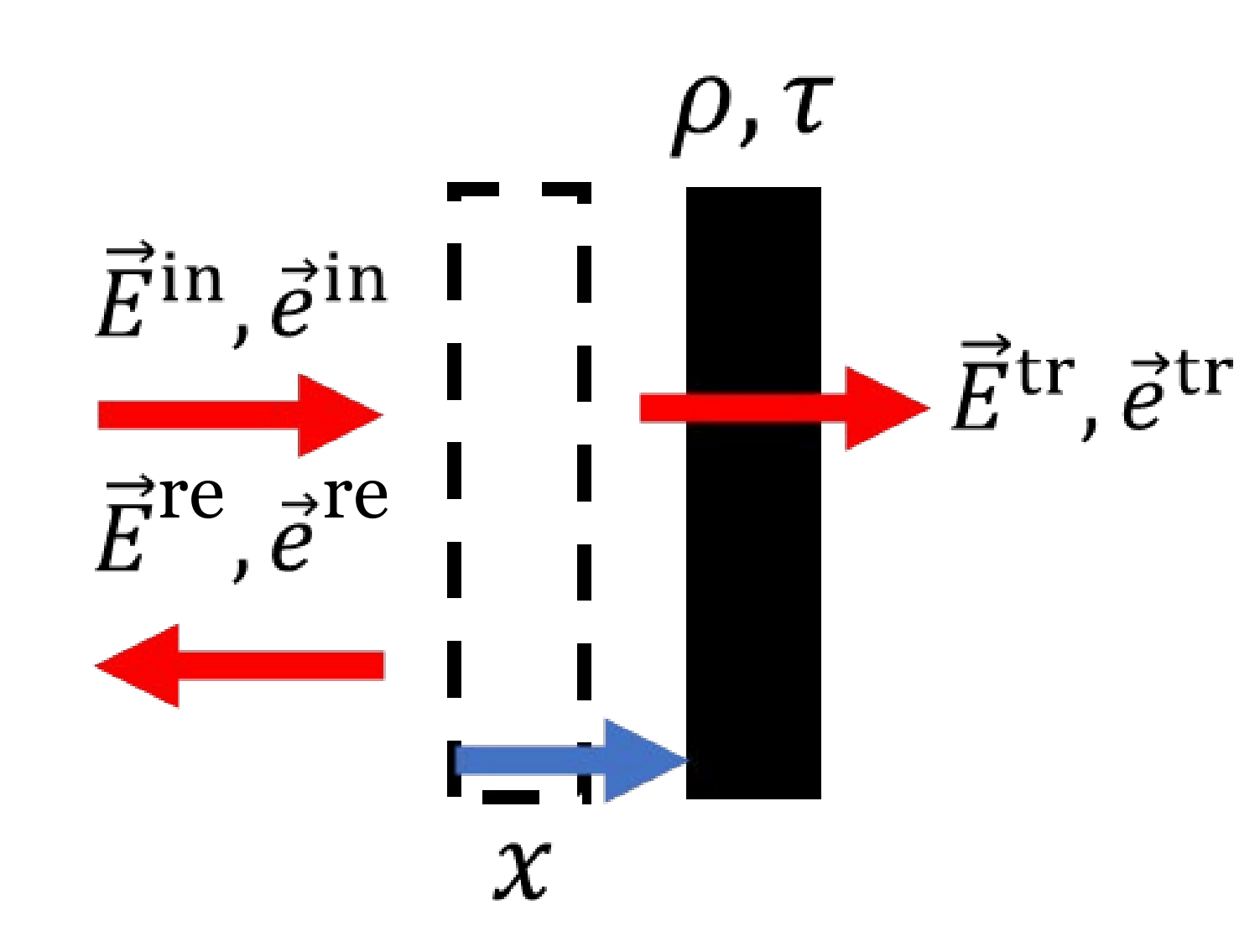}
    \end{minipage}
    \caption{Schematic diagram of quadrature component vectors for (left) a free space element of length $L$ and (right) a moving mirror with amplitude reflectivity and transmissivity coefficients $\rho$ and $\tau$ which is displaced from its mean position by a distance $x\ll\lambda$.}
    \label{fig:free space and moving mirror}
\end{figure}

\subsubsection{Free space}
\label{sec: free space operators}
The electric fields at two ends of a length of vacuum $L$ (left half of Fig. \ref{fig:free space and moving mirror}) are related by a simple time delay of $L/c$.
Applying this delay in the frequency domain gives
\begin{equation}
\begin{aligned}
    \vec E^\text{out}&=\mathbb{R}(kL)\vec E^\text{in},
    \\
    \vec e^\text{ out}(\Omega)&=\exp(i\Omega L/c)\mathbb{R}(kL)\vec e^\text{ in},
    \end{aligned}
    \label{eq: free space propagation}
\end{equation}
where $k=\omega_0/c$ and
\begin{equation}
    \mathbb{R}(\varphi)\equiv\begin{pmatrix}
        \cos\varphi & -\sin\varphi \\ \sin\varphi & \cos\varphi
    \end{pmatrix}.
\end{equation}
Across free space, both the carrier and sideband quadrature fields experience a rotation, with the sideband fields also picking up a phase shift due to the time delay. 

\subsubsection{Moving mirror}
For a moving mirror, we fix the points where the incident, reflected, and transmitted fields (right half of Fig. \ref{fig:free space and moving mirror}) are evaluated to either side of the infinitesimally-thin mirror at its mean position. With this convention, the transmitted field through a mirror of amplitude reflectivity $\rho>0$ and transmissivity $\tau$ is simply
\begin{equation}
    \mathcal{E}^\text{tr}(t)=\tau\mathcal{E}^\text{in}(t),
\end{equation}
with $\mathcal{E}^\text{tr}$ being the transmitted portion of the full incident field $\mathcal{E}^\text{in}$. The quadrature vectors thus satisfy
\begin{equation}
    \vec E^\text{tr}=\tau\vec E^\text{in},\quad \vec e^\text{ tr}(\Omega)=\tau\vec e^\text{ in}(\Omega).
\end{equation}
Meanwhile, the full reflected field is given by
\begin{equation}
    \mathcal{E}^\text{re}(t)=\mathcal{E}^\text{in}(t-2x(t)/c),
\end{equation}
where $x(t)$ is the perturbative displacement of the moving mirror from its equilibrium position, with positive values indicating a displacement away from the source of the incident field. To first order in $kx$, this results in
\begin{equation}
    \vec E^\text{re}=-\rho\vec E^\text{in},~ \vec e^\text{ re}(\Omega)=-\rho(\vec e^\text{ in}(\Omega)-2kx(\Omega)\vec E^\text{in}_\perp),
    \label{eq: moving mirror reflected fields}
\end{equation}
where $\vec E^\text{in}_\perp\equiv\begin{pmatrix} E_2^\text{in} \\ -E_1^\text{in} \end{pmatrix}=\mathbb{R}(-\frac\pi2)\vec E^\text{in}$. The sidebands thus pick up perturbative contributions from the carrier field.

\subsubsection{Moving dielectric stack}
\label{sec: moving dielectric}
Handling a moving dielectric stack (which will serve as the model for the plate-like sensor) is a bit more subtle. In principle, one could construct multilayer dielectric sensors with spatially varying indices of refraction. Such multilayer dielectrics are in fact well-motivated for improved detector sensitivity \cite{LSD2020}; furthermore, it is straightforward to generalize the formulae in this section to multilayer sensors. That being said, we find that even a single material dielectric is sufficient to reveal a wealth of interesting phenomena in the cavity-sensor system. We restrain ourselves to a single layer dielectric slab throughout the remainder of this work.

\begin{figure}
    \centering
    \includegraphics[width=\linewidth]{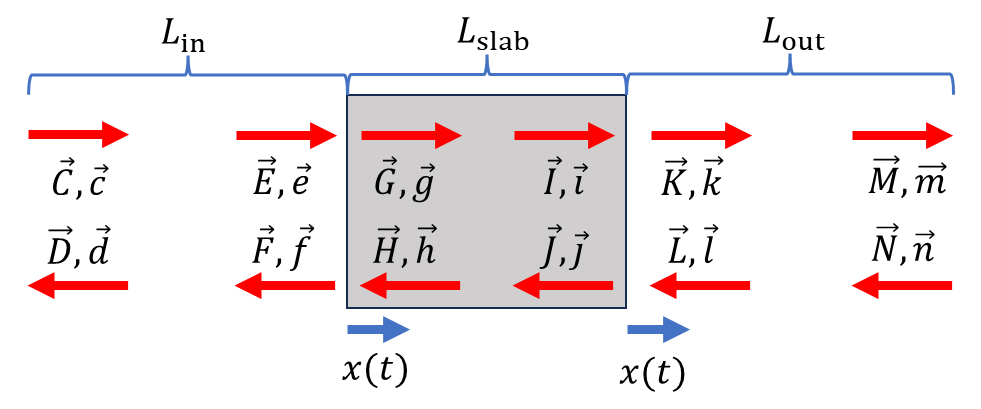}
    \caption{Schematic diagram of the carrier and sideband fields entering, propagating within, and leaving a moving dielectric slab (gray) of length $L_s$ which experiences displacement $x(t)$. Field vector labels are chosen for comparison to Fig. \ref{fig:fields diagram complete}.}
    \label{fig: moving dielectric slab}
\end{figure}

Consider the schematic diagram in Fig. \ref{fig: moving dielectric slab}. At first glance, it seems as though the slab can be handled by treating each face of the slab as a moving mirror and treating the propagation from $(\vec G,\vec g)$ to $(\vec I,\vec i)$ and from $(\vec J,\vec j)$ to $(\vec H,\vec h)$ as free space elements, accounting for the reduced speed of light in the dielectric, $c/n$, with $n$ being its index of refraction. 

However, fixing the location of the field evaluations in this way requires an additional correction. Consider the propagation from $\vec e$ to $\vec g$ through the vacuum-dielectric interface when the slab moves. When the slab moves to the right ($x(t)>0$), the path length from $\vec e$ to a fixed coordinate within the slab gains $x(t)$ worth of vacuum and loses $x(t)$ worth of dielectric. The resulting phase difference must be accounted for in a similar way to the reflection off of a moving mirror. We insert this contribution into the sideband fields generated during the transmission through the vacuum-dielectric interface:

\begin{subequations}
\begin{alignat}{2}
    \vec f &= -\rho_s\bigl(\vec e-2kx\vec E_\perp\bigr)
        +\tau_s^\text{out}\bigl(\vec h-k(1-n)x\vec H_\perp\bigr), \\
    \vec g &= \tau_s^\text{in}\bigl(\vec e+k(n-1)x\vec E_\perp\bigr)
        +\rho_s\bigl(\vec h+2nkx\vec H_\perp\bigr), \\
    \vec h &= \mathbb{R}(knL_s)\exp(i\Omega nL_s/c)\vec j, \\
    \vec i &= \mathbb{R}(knL_s)\exp(i\Omega nL_s/c)\vec g, \\
    \vec j &= \rho_s\bigl(\vec i-2nkx\vec I_\perp\bigr)
        +\tau_s^\text{in}\bigl(\vec l-k(n-1)x\vec L_\perp\bigr), \\
    \vec k &= \tau_s^\text{out}\bigl(\vec i+k(1-n)x\vec I_\perp\bigr)
        -\rho_s\bigl(\vec l+2kx\vec L_\perp\bigr).
\end{alignat}
    \label{eq: moving dielectric sideband fields, old convention}
\end{subequations}

\begin{figure*}
    \centering
    \includegraphics[width=0.8\linewidth]{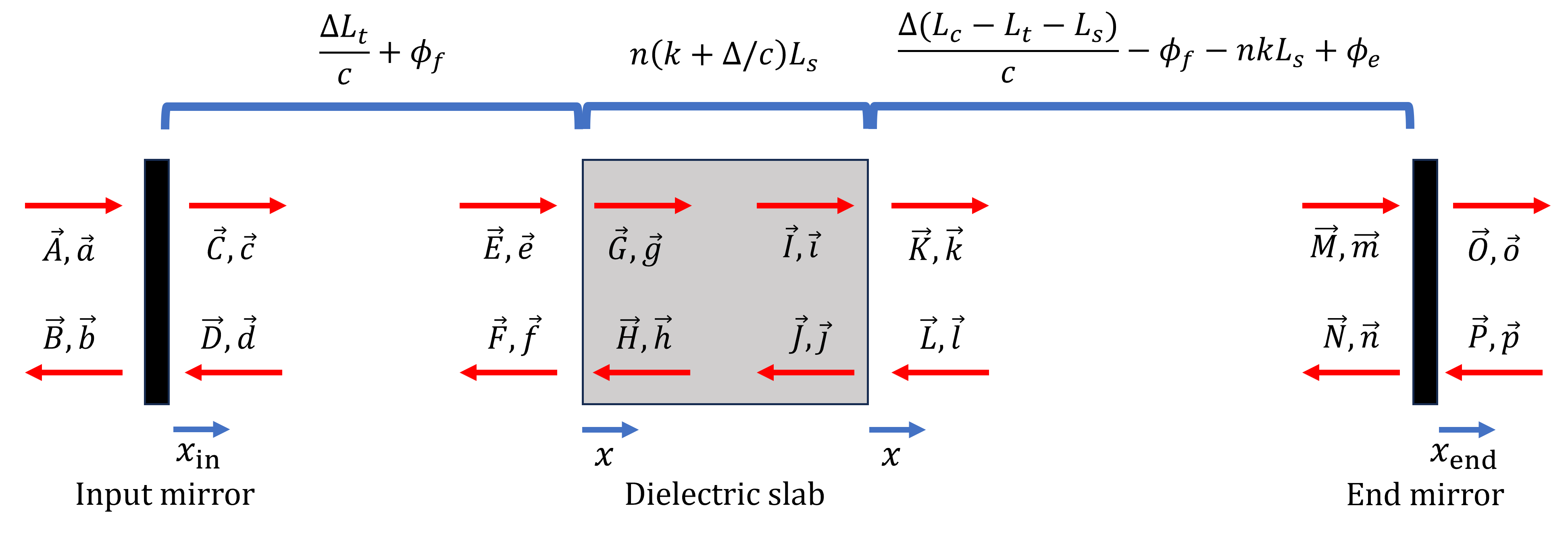}
    \caption{Diagram of the carrier and sideband fields in the cavity with a levitated dielectric. The labels for vectors describing the fields at various locations will be used in a number of calculations within Sec. \ref{sec: cavity equations and optomechanical loops}. The relative phase accumulated over free propagation elements for a laser with detuning $\Delta$ is written above the blue brackets.}
    \label{fig:fields diagram complete}
\end{figure*}

Here, we have suppressed the $\Omega$ dependence in the sideband fields and in $x$, and $L_s$ is the thickness of the slab. The ingoing and outgoing transmissivities and the reflectivity for the vacuum-dielectric interface are \cite{Byrnes2016}
\begin{equation}
    \tau_s^\text{in}=\frac{2}{n+1},\quad \tau_s^\text{out}=\frac{2n}{n+1},\quad\rho_s=\frac{n-1}{n+1}.
\end{equation}
To facilitate comparison with the full system model (see Fig. \ref{fig:fields diagram complete}), we have adopted identical field vector labels; however, the lengths and phases in the full model are defined by experimental parameters. The corresponding equations for the DC field vectors can be found by taking $\Omega$ and $x(\Omega)$ to zero and replacing all lowercase letter vectors with their uppercase counterparts.

However, we can modify the field conventions slightly in a way which reduces the mathematical complexity of the moving slab. If we instead fix the location of the field evaluations $\vec E$ through $\vec L$ (and their sideband counterparts) to the surfaces of the slab \textit{as it moves}, the space between $\vec G$ and $\vec I$, etc. is always completely dielectric.
In order to accomplish this within the current framework, we must extend the ``moving dielectric slab" element to also include the free space elements on either side of the dielectric (of nominal lengths  $L^\mathrm{in}$ and $L^\mathrm{out}$, respectively). These lengths are arbitrary, but for our system, they will be the distances from the input and end mirrors to the slab surfaces, respectively. As the dielectric moves, the length of these elements changes dynamically, thereby modifying the phase imprinted on the sideband fields at the vacuum-dielectric interfaces. Under this convention, the propagation equations are as follows:

\begin{subequations}
\begin{alignat}{2}
    \vec d &= \mathbb{R}(kL^\text{in})\bigl(\exp(i\Omega L^\text{in}/c)\vec f - kx\vec F_\perp\bigr), \\
    \vec e &= \mathbb{R}(kL^\text{in})\bigl(\exp(i\Omega L^\text{in}/c)\vec c - kx\vec C_\perp\bigr), \\
    \vec f &= -\rho_s\vec e + \tau_s^\text{out}\vec h, \\
    \vec g &= \tau_s^\text{in}\vec e + \rho_s\vec h, \\
    \vec h &= \mathbb{R}(knL_s)\exp(i\Omega nL_s/c)\vec j, \\
    \vec i &= \mathbb{R}(knL_s)\exp(i\Omega nL_s/c)\vec g, \\
    \vec j &= \rho_s\vec i + \tau_s^\text{in}\vec l, \\
    \vec k &= \tau_s^\text{out}\vec i - \rho_s\vec l, \\
    \vec l &= \mathbb{R}(kL^\text{out})\bigl(\exp(i\Omega L^\text{out}/c)\vec n + kx\vec N_\perp\bigr), \\
    \vec m &= \mathbb{R}(kL^\text{out})\bigl(\exp(i\Omega L^\text{out}/c)\vec k + kx\vec K_\perp\bigr).
\end{alignat}
\label{eq: moving dielectric sideband fields}
\end{subequations}

We have verified that both the modified convention which generates Eq. \eqref{eq: moving dielectric sideband fields} and the adaptation of the standard moving mirror elements to the dielectric-vacuum interfaces (Eq. \eqref{eq: moving dielectric sideband fields, old convention}) produce identical observable results --- namely, the outgoing fields $\vec d$ and $\vec n$ in terms of $\vec c$, $\vec m$, and $x$, as well as the DC and AC forces on the slab. Thus, both conventions are sufficient to model the dielectric sensor; for simplicity, we elect to use Eq. \eqref{eq: moving dielectric sideband fields}.

We now have a complete set of tools to write down the system of equations describing the fields in the cavity with a levitated dielectric. Through this formalism, all perturbations to the system --- motion of the sensor, small displacements of the cavity mirror surfaces, and fluctuations in the input light fields --- will produce sideband fields which are \textit{linear} in the size of the initial perturbation. These relations completely determine the optical fields for a specified sensor position; the remaining ingredient needed for a dynamical response is the force exerted by those fields on the sensor.

\subsection{Optical forces from carrier and sideband fields}
\label{sec: sideband field forces}

Self-consistent calculations of the experimental response to some external input require computing the force on the sensor from optical sideband fields. These forces drive the sensor's motion, which is in turn imprinted on the readout via the generation of additional sideband fields \footnote{We assume that the positions of the cavity mirrors are insensitive to optical forces, but can still be displaced via coating thermal noise, etc.}. Furthermore, the forces from the DC (carrier) fields also play a crucial role, setting the equilibrium position around which the sensor oscillates.

Standard treatments of levitated dielectrics in the Rayleigh regime (size $\lesssim\lambda$) operate from a perspective where the total force on the sensor is separated into two pieces. These two pieces are often labeled as radiation pressure or a ``scattering force", sourced by direct momentum transfer from the electromagnetic field, and the ``gradient force", which arises from position-dependent variations in the stored field momentum within the sensor. Approximating the small dielectric as a single electric dipole and computing the Lorentz force reveals that the gradient force aptly scales with the gradient of the local field intensity \cite{Gordon, GordonplusAshkin, Ashkin:1978zsu}. It is the balance of these two forces which results in Rayleigh-regime dielectric objects trapped within Gaussian laser beams to lie at the center of the beam's cross section but slightly downfield from the focus \cite{Ashkin:86, ChangSpheres}.

The geometry of our class of levitated sensors, however, lies well outside the Rayleigh regime, with the thickness being comparable to or larger than $\lambda$ and the radius well exceeding $\lambda$. With this in mind, the configuration of this experiment is more akin to a membrane-in-the-middle or membrane-at-the-end optomechanical setup. This observation suggests that the usual Rayleigh-regime gradient-plus-scattering decomposition is not sufficient to describe the force in this geometry. Instead, the force must be computed in a more generic way, e.g., by integrating the Maxwell stress-energy tensor over a bounding surface enclosing the sensor's volume. Fortunately, the planar nature of the sensor's surfaces along the cavity axis makes this computation relatively straightforward to handle analytically.

In a companion paper \cite{ForceLetter}, we carry out a complete derivation of the force on a plate-like levitated sensor within the Fabry-Pérot cavity\footnote{Pedantically, by force on the sensor we mean ``time derivative of its center of mass momentum". There are other momenta (e.g., the field momentum) whose time-derivative differs from these results.}. Leaving the details to that work, the DC force on the sensor is  
\begin{equation}
    F_\text{DC}=\frac12\epsilon_0A_\text{eff}(\vec E\cdot\vec E+\vec F\cdot\vec F-\vec K\cdot\vec K-\vec L\cdot\vec L),
    \label{eq: DC force}
\end{equation}
with $A_\text{eff}$ being an effective beam area, given by
\begin{equation}
    A_\text{eff}\equiv\frac{\pi w_0^2}{2}\Bigr(1-e^{-2R^2/w_0^2}\Bigr).
\end{equation}
Inserting parameters from Table \ref{tab:parameters} into this expression, the total captured power is $(1-e^{-8})=99.97\%$ of the total beam power, justifying our plane wave model. Meanwhile, the AC force is
\begin{multline}
    \delta F(\Omega)=\epsilon_0A_\text{eff}(\vec E\cdot\vec e+\vec F\cdot\vec f-\vec G\cdot\vec g-\vec H\cdot\vec h
    \\+\vec I\cdot\vec i+\vec J\cdot\vec j-\vec K\cdot\vec k-\vec L\cdot\vec l).
    \label{eq: AC force}
\end{multline}
Here quadrature vector labels coincide with those used in Fig. \ref{fig: moving dielectric slab} and $\Omega-$dependence in the sideband fields has been suppressed. This is the only instance in which we will account for the Gaussian cross-sectional profile of the beam. Note that the DC force also implicitly includes a contribution scaling with $-\vec G\cdot\vec G-\vec H\cdot\vec H+\vec I\cdot\vec I+\vec J\cdot\vec J$, but this vanishes by conservation of momentum ($\vec H$ and $\vec I$ are just rotations of $\vec J$ and $\vec G$, respectively). We will see in Sec. \ref{sec: closing optomechanical loops and transfer functions} that the AC force, being linear in the sideband field vectors, is immediately amenable to closing optomechanical loops in the transfer functions.

\subsection{Details of the experimental model}
In Fig. \ref{fig:fields diagram complete}, we show the full set of carrier and sideband fields in our model, and in Appendix \ref{app: full field equations}, we write down the full system of optical equations relating the fields to one another. Small perturbations (i.e., inputs for transfer functions) can enter the system through seven ports: four from the optical field fluctations entering through input and end mirrors ($a_1$, $a_2$, $p_1$, $p_2$), two from mirror displacements ($x_\text{in}$, $x_\text{end}$), and one from external forces on the sensor ($F_\text{ext}$).

We broadly assume that macroscopic lengths $L_c$ and $L_t$ are integer multiples of the trap laser wavelength, which is defined by $\Delta=0$. DC phase differences (modulo $2\pi$) for fields at either ends of these free space elements are encoded through $\Delta\ll\omega_0$ and by assigning small phase angles to interfaces (such as $\phi_f$) to encode microscopic length changes. The expressions for these DC phase differences appear above the blue brackets highlighting these free space elements. 

The microscopic position of the end mirror is captured by the parameter $\phi_e$, defined as half of the phase, modulo $2\pi$, accumulated by the trap laser in one round trip of the full cavity. In other words, the DC phase differences in each segment of the cavity sum to $\phi_e$ when $\Delta=0$. Establishing a convention for setting $\phi_e$ physically amounts to choosing a procedure for ``locking" the cavity. 

In this work, we define a ``locked" cavity as one where the transmitted power through the end mirror is maximized over the cavity length. In practice, a Pound-Drever-Hall locking procedure is employed to keep the length slightly off resonance, where the response to length fluctuations is linear rather than quadratic \cite{PDHlocking}. However, to minimize the number of free parameters in this work, we assume the true maximum can be attained. To locate the parameters of the ``lock", we iteratively swap between maximizing the transmitted power over $\phi_{e}$ and computing the new trap location $\phi_{f,\text{trap}}$ until both values find an equilibrium within a predetermined tolerance. In other words, we simultaneously solve
\begin{equation}
    F_\text{DC}=0,\quad dF_\text{DC}/d\phi_f<0,\quad dP_\text{trans}/d\phi_e=0.
\end{equation}
Up to trivial additions of $N\times\pi$ to the phase angles, our iterative procedure identifies a single solution ($\phi_{f,\text{lock}}, \phi_{e,\text{lock}}$) for all $L_s\not=N\times\lambda/2n$. We take the locking bandwidth frequency to be much smaller than the resonant frequency of the sensor's motion so that the locking mechanism can be ignored when computing transfer functions within the resonant band.

\subsection{System response including optomechanical back-action}
\label{sec: closing optomechanical loops and transfer functions}
From the discussions above, it is clear that \textit{(i)} the motion of the sensor generates sideband fields, and \textit{(ii)} sideband fields produce forces on the sensor, which cause it to move. In order to calculate fully self-consistent transfer functions, we must simultaneously solve for sensor motion and sideband field generation via inclusion of the optical back-action terms.

The first step in achieving this is assigning a mechanical susceptibility $\chi_\text{mech}(\Omega)$ to the sensor so that $x(\Omega)=\chi_\text{mech}(\Omega)\delta F(\Omega)$. When computing sensitivity estimates for measurements on the probe beam, we only ever need to compute transfer functions with back-action when studying the response in the probe fields. So, we can determine the parameters of the mechanical susceptibility just from properties of the environment and the trap laser. The resonant frequency of the sensor's motion is set by the DC forces from the trap laser:
\begin{equation}
    \Omega_\text{trap}^2=-\frac{k^2}{m}\frac{dF_\text{DC,trap}}{d\phi_f}\Bigr|_{\phi_{f,0}},
\end{equation}
where $m$ is the sensor mass, $F_\text{DC,trap}$ is the DC force from the trap beam, and $\phi_{f,0}$ is a value of $\phi_f$ where $F_\text{DC,trap}=0$. Meanwhile, the mechanical damping rate is set by the rate of collisions between the sensor and gas molecules in the environment \cite{LSD2020},
\begin{equation}
    \Gamma_\text{gas}=32PR^2/\bar vm,
    \label{eq: gas damping rate}
\end{equation}
where $P$ is the gas pressure and $\bar v\equiv \sqrt{3k_BT/m_\text{gas}}$ is the rms velocity of a gas molecule in the environment, with $k_B$ being Boltzmann's constant. For our experiment, we choose $P=10^{-9}$ torr, $T=300$K, and estimate $m_\text{gas}\approx28$ amu (the mass of an N${}_2$ molecule). The mechanical susceptibility is constructed from $\Omega_\text{trap}$ and $\Gamma_\text{gas}$ as follows:
\begin{equation}
    \chi_\text{mech}(\Omega)=\Bigr(m(\Omega_\text{trap}^2-\Omega^2-i\Gamma_\text{gas}\Omega)\Bigr)^{-1}.
    \label{eq: chi mech}
\end{equation}

In full generality, the susceptibility function must also account for probe laser contributions to the resonant frequency and damping. However, since the probe laser power is far smaller than the trap laser power, these contributions are relatively weak --- we will see in Sec. \ref{sec: probe beam cooling} that they can be interpreted as a shift to the poles of $\chi_\text{mech}$ located at the resonant frequencies.

The sideband fields are already expressed in linear terms of $x$ and these perturbations, so all that remains is to ensure that $x$ is also written in linear terms of the sidebands and/or perturbations. This is already satisfied trivially, as the force given in Eq. \eqref{eq: AC force} is expressed directly in linear terms of the sideband fields. Therefore, the AC system of equations can incorporate back-action physics by adding one final equation to system which relates the intracavity field amplitudes to one another:
\begin{equation}
    x(\Omega)=\chi_\text{mech}(\Omega)\Bigl(\delta F_\text{ext}(\Omega)+\delta F(\Omega)\Bigr).
    \label{eq: force coefficient sum}
\end{equation}
While fluctuations in the trap fields do not couple directly into the readout of the probe beam, they do source forces on the sensor, generating displacements which create sidebands of the probe field. So in the expression above, $\delta F(\Omega)$ must include all contributions to the probe and trap beam optical forces. That is, $\delta  F(\Omega)=\delta F_\text{probe}(\Omega)+\delta F_\text{trap}(\Omega)$, where
\begin{equation}
    \delta F_\text{probe}(\Omega)=\sum_{\substack{z\in\{x,~x_\text{in},~x_\text{end},\\a_1,~a_2,~p_1,~p_2\}}}z\times \frac{\partial}{\partial z}\delta F(\Omega),
    \label{eq: probe beam differential force}
\end{equation}
where the partial derivatives here are evaluated for the probe laser properties and $a_1$, $a_2$, $p_1$, and $p_2$ are fluctuations in the incoming probe fields, and 
\begin{equation}
    \delta F_\text{trap}(\Omega)=\sum_{\substack{z\in\{x_\text{in},~x_\text{end},\\a_1,~a_2,~p_1,~p_2\}}}z\times\frac{\partial }{\partial z}\delta F(\Omega),
    \label{eq: trap beam differential force}
\end{equation}
where the partial derivatives and $a_1$, etc. refer to the trap beam fields. Note the absence of $x$ in the summation for the trap beam force --- this partial derivative is already incorporated through $\chi_\text{mech}$.

As a result, solving the DC cavity field equations reduces to solving a 28-equation linear system\footnote{$A_1$ and $A_2$ are determined by the laser properties, and we set $P_1=P_2=0$ since no laser enters from the right.} ($B_{1,2}$ through $O_{1,2}$), and solving the AC equations reduces to solving a 29-equation linear system ($b_{1,2}$ through $o_{1,2}$, plus $x$), which we do by numerically inverting the corresponding square matrices.

With back-action included, the system is still fully linear in its inputs, so the detector response to a perturbation which enters coherently across multiple input ports can be computed via summation over each input. This feature is useful for more complicated optical layouts, such as a Fabry-Pérot Michelson interferometer configuration, where an incident GW appears in multiple ports by stretching one arm while shrinking the other.

\subsection{Cooling from probe beam}
\label{sec: probe beam cooling}
In addition to carrying signatures of perturbations to the experiment, the probe beam also modifies the mechanical susceptibility through the usual dynamical back-action. In recent years, optical cavities have been employed to achieve passive cooling of micro-mechanical systems, both through radiation pressure from detuned light fields \cite{ChangSpheres,ChangThinDiscs,KieselCooling,FonsecaCooling} and through preferential coherent scattering into a particular cavity mode \cite{Delic:2019gri,Delic:2019xqd,Windey:2019yoi,Pontin:2022ywq}. In this section, we demonstrate how to compute the cooling effect of the detuned probe on the levitated sensor's motion along the cavity axis. 

The mechanical susceptibility $\chi_\text{mech}$ (due to the trap laser) features a pole at
\begin{equation}
    \Omega_p=\sqrt{\Omega_\text{trap}^2-\frac14\Gamma_\text{gas}^2}-\frac12i\Gamma_\text{gas}\approx \Omega_\text{trap}-\frac12i\Gamma_\text{gas}.
\end{equation}
When the probe beam is added in, we can rewrite Eq. \eqref{eq: force coefficient sum} in the form
\begin{multline}
    x(\Omega)\Bigr(\chi_\text{mech}^{-1}(\Omega)+K(\Omega)\Bigr)
    \\
    \equiv x(\Omega)\chi_\text{cooled}^{-1}(\Omega)=\delta F_\text{tot}(\Omega),
\end{multline}
where 
\begin{equation}
    K(\Omega)\equiv -\frac{\partial}{\partial x(\Omega)}\delta F_\text{probe}(\Omega).
\end{equation} and $\delta F_\text{tot}(\Omega)=\delta F_\text{ext}(\Omega)+\delta F(\Omega)$ includes all differential forces on the sensor due to incident perturbations (mirror displacements, laser fluctuations, etc.).

The inclusion of $K(\Omega)$ shifts the resonant frequency and damping linewidth of the sensor in the trap --- this is a manifestation of the optical spring effect observed in \cite{AggarwalThesis}. Assuming the probe beam forces are small compared to the trap beam forces, we can find the frequency and linewidth shifts perturbatively. To do so, we assume that $K(\Omega)$ is analytic in a neighborhood of the unperturbed pole and contains no nearby optical pole. For our setup, $K$ evolves across the real frequency axis on the scales of the cavity linewidth $\kappa$. So, as long as $\Gamma_\text{gas}\ll\kappa$ and the shift in the pole frequency $\delta\Omega_p$ satisfies $\mathrm{Re}(\delta\Omega_p)\ll\kappa$ and $\mathrm{Im}(\delta\Omega_p)\ll\kappa$ (both of which are valid assumptions for typical cavity and mechanical configurations), the Cauchy-Riemann relations \cite{Bak2010} imply that an expansion around $\Omega_p$ to linear order in $\delta\Omega_p$ is sufficiently accurate.

The zero of $\chi_\text{cooled}^{-1}$ thus appears at $\Omega_p+\delta\Omega_p$, where $\delta\Omega_p$ satisfies
\begin{multline}
    m(-2\Omega_p\delta\Omega_p-i\Gamma_\text{gas}\delta\Omega_p)
    \\
    +\Bigr(K(\Omega_\text{trap})+(-\frac12i\Gamma_\text{gas}+\delta\Omega_p) K'(\Omega_\text{trap})\Bigr)=0.
\end{multline}
After algebraic manipulation, this yields
\begin{equation}
    \delta\Omega_p=\frac{K(\Omega_\text{trap})-\frac12i\Gamma_\text{gas}K'(\Omega_\text{trap})}{m(2\Omega_\text{p}+i\Gamma_\text{gas})-K'(\Omega_\text{trap})}.
\end{equation}
Since $\kappa\gg\Omega_\text{trap}\gg \Gamma_\text{gas}$, this is well approximated by
\begin{equation}
    \delta\Omega_p\approx \frac{K(\Omega_\text{trap})}{2m\Omega_\text{trap}}\Bigr(1+\frac{K'(\Omega_\text{trap})}{2m\Omega_\text{trap}}\Bigr).
    \label{eq: delta Omega p approx}
\end{equation}
Finally, from the definition of $\Omega_p$, we find that the parameters defining $\chi_\text{cooled}$ are
\begin{subequations}
\begin{equation}
    \Omega_\text{cooled}=\Omega_\text{trap}+\text{Re}(\delta\Omega_p),
\end{equation}
\begin{equation}
    \Gamma_\text{cooled}\equiv\Gamma_\text{gas}+\Gamma_\text{probe}= \Gamma_\text{gas}-2\text{Im}(\delta\Omega_p).
\end{equation}
\end{subequations}
These results match numerical tests on transfer functions at the sub-percent level. 
The sign of $\text{Im}(\delta\Omega_p)$ determines whether the motion along the cavity axis is heated or cooled by the probe beam, with $\text{Im}(\delta\Omega_p)<0$ corresponding to increased damping.

We emphasize that $\chi_\text{cooled}$ should \textit{not} be substituted for $\chi_\text{mech}$ in Eq. \eqref{eq: force coefficient sum} in the presence of the probe. Rather, $\chi_\text{cooled}$ should be thought of as an effective susceptibility which describes the closed-loop displacement response of the sensor to an oscillatory force -- that is, the frequency-dependent optical back-action $\delta F(\Omega)$ has been captured by the new effective resonant frequency and damping linewidth.

\subsection{Noise budgets from the two-photon formalism}
\label{sec: noise budget construction}

Having demonstrated how to compute self-consistent transfer functions, we can finally proceed to the calculation of the strain-normalized noise appearing in the detector (i.e., the frequency-domain sensitivity). While we present the mathematical steps for this procedure here for completion's sake, we will defer the explicit definition of noise spectral densities and the analysis of numerical sensitivity estimates to a future paper.

Computing the estimated strain sensitivity requires projecting each noise source onto the readout channel of choice. For concreteness, we will assume a quadrature measurement on the reflected sideband field $\vec b$ with quadrature angle $\xi$, but other measurement channels can be substituted straightforwardly. For this particular measurement scheme, $S_{\xi,N_i}$, the PSD for the noise in the readout channel from noise source $N_i$, is \cite{CorbittFramework}
\begin{equation}
    S_{\xi,N_i}= \begin{pmatrix}
    \cos\xi & \sin\xi \end{pmatrix}T_{b,n_i}S_{N_i}T_{b,n_i}^\dagger
    \begin{pmatrix}
    \cos\xi \\ \sin\xi
    \end{pmatrix},
\end{equation}
where $S_{N_i}$ is the PSD of the given noise source and $T_{b,n_i}$ is the transfer function matrix from the port where that noise source enters the experiment $(n_i)$ to the $\vec b$ sideband quadrature components.

With all the noise sources projected onto the quadrature readout channel, we then convert these projected noise PSDs into units of strain. To do so, we simply normalize with the GW response function, given by
\begin{equation}
    T_{b,h}=
    \begin{pmatrix}
        \cos\xi & \sin\xi
    \end{pmatrix}
    \begin{pmatrix}
        \partial b_1/\partial h \\
        \partial b_2/\partial h
    \end{pmatrix}.
\end{equation}
The strain-normalized PSD for each source is just
\begin{equation}
    S_{hh,N_i}=S_{b,N_i}/|T_{b,h}|^2,
\end{equation}
and the total PSD for noise in strain is the sum of the individual PSDs, i.e.
\begin{equation}
    S_{hh}=\sum_{N_i}S_{hh,N_i}.
\end{equation}

The structure of $T_{b,n_i}$ contains a subtlety that is worth mentioning. For noise sources with a scalar PSD (e.g., gas damping force noise, mirror substrate thermal noise), the transfer function matrix is
\begin{equation}
    T_{b,n_i}\equiv 
    \begin{pmatrix}
        \partial b_1/\partial n_i \\
        \partial b_2/\partial n_i
    \end{pmatrix},
\end{equation}
where the partial derivatives indicate the transfer function from port $n_i$ to each quadrature component of the reflected sideband field. Meanwhile, noise sources which have a $2\times2$ matrix PSD (e.g., quantum noise from the trap and probe beams) admit a transfer function matrix of the form
\begin{equation}
    T_{b,n_i}\equiv 
    \begin{pmatrix}
        \partial b_1/\partial n_{i,1} & \partial b_1/\partial n_{i,2} \\
        \partial b_2/\partial n_{i,1} & \partial b_2/\partial n_{i,2}
    \end{pmatrix},
\end{equation}
where the $1$ and $2$ subscripts indicate the quadrature component of the fields where the noise is introduced.

Naively, one might anticipate that greater cooling of the thermal motion of the sensor would lead to a proportional improvement in the detector sensitivity. However, because the cooling also influences the magnitude of the response to an incident gravitational wave, for example, this relation does not hold generically. The precise role of cooling in the complete strain-normalized PSD depends on the relative contributions from noise sources which propagate to the readout through distinct channels, a concept we explore with an illustrative noise profile in Appendix \ref{app: noise vs Q}. 

\section{Detector response}
\label{sec: detector response}

The formalism above can be used to calculate the response of the coupled cavity-sensor system to external perturbations. Here we focus on transfer functions from external forces, laser field fluctuations, and cavity mirror displacements to the reflected probe field, together with the gravitational-wave response.
\subsection{Transfer functions and GW response}

\begin{figure}
    \centering
    \includegraphics[width=\linewidth]{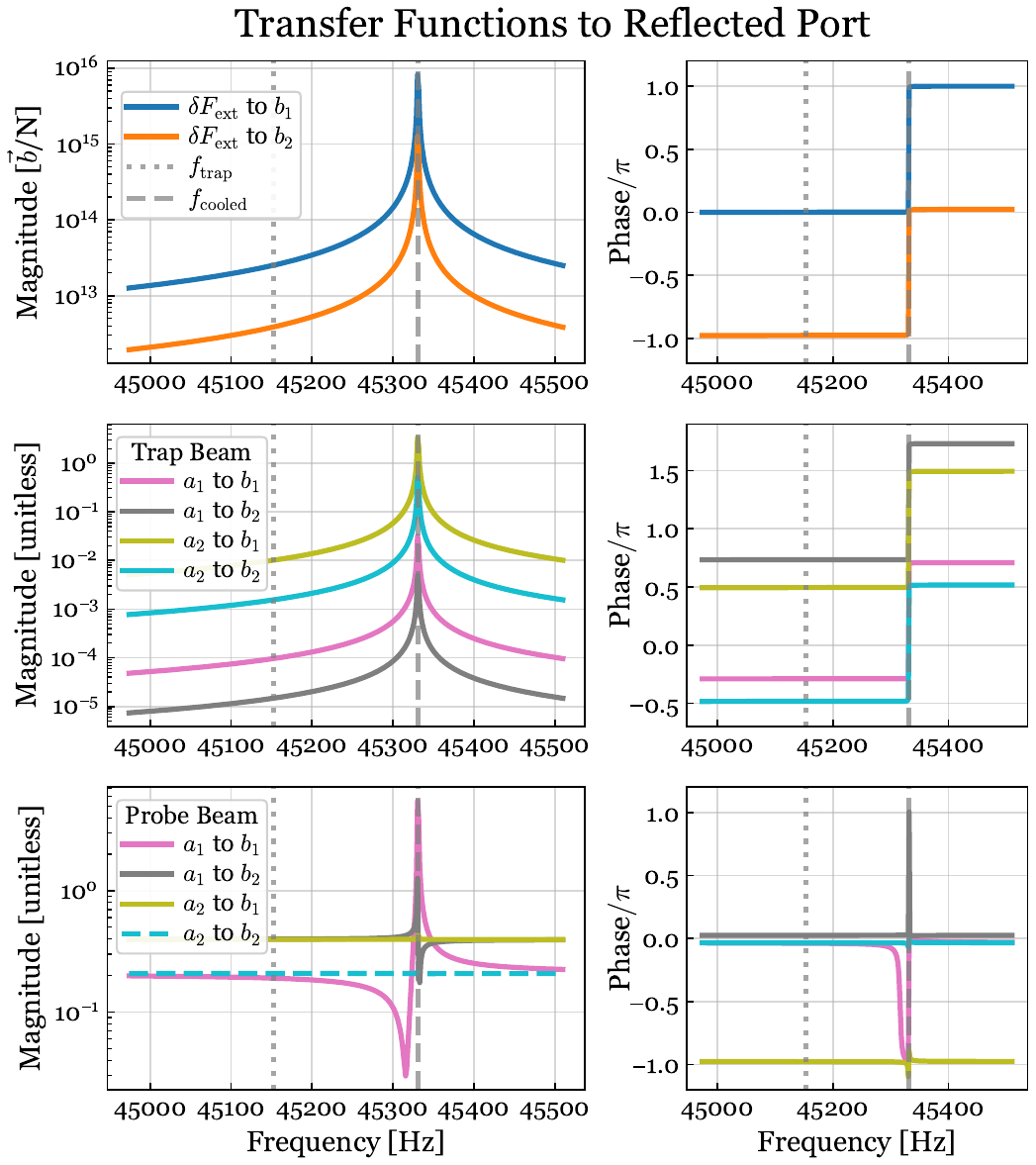}
    \caption{A set of transfer functions to the two quadratures of the reflected readout port. Here, we set the parameters $L_s=0.25\lambda/n$ and $\Delta=-0.5\kappa$. Gray lines mark the trap resonant frequency (dotted) and new resonant frequency after including the probe beam's optical spring effect (dashed). Transfer function input ports include an external force (top), trap beam field fluctuations (middle), and probe beam field fluctuations (bottom). For the laser fluctuation entries, we only plot the response to $a_1$ and $a_2$ here because their analogues in $p_1$ and $p_2$ do not provide any noteworthy additional information.}
    \label{fig: example transfer functions}
\end{figure}

We now turn to the overall response of the detector to various incident perturbations. In Figs. \ref{fig: example transfer functions} and \ref{fig: example mirror disp transfer functions}, we plot a set of example transfer functions from a variety of input ports --- external forces, laser field fluctuations, and mirror displacements --- to the quadrature components of the field at the readout port $\vec b$ (measuring fields which leave through the cavity input mirror). We will now highlight certain features within each class of transfer functions which offer insight into the response of the experiment to general stimuli.

\subsubsection{Response to external forces}

Perhaps the simplest structure is that of the transfer functions from an external force on the sensor to the readout (top panels of Fig. \ref{fig: example transfer functions}). We identify a peak in these transfer functions at the resonant frequency accounting for the probe's optical spring effect, $f_\text{cooled}=(\Omega_\text{trap}+\mathrm{Re}(\delta\Omega_p))/2\pi$, consistent with the shift of the pole of the mechanical response of the sensor. 

Furthermore, it is evident that the response of the system to a force on the sensor is much stronger in the amplitude quadrature ($b_1$) than the phase quadrature ($b_2$). For a cavity held on resonance with its probe field, the response of a moving mirror generally appears in the phase quadrature. However, because the probe beam is detuned here, the intensity in the cavity is far more sensitive to the motion of the dielectric than the optical resonant frequency is. This renders the coupling to the readout to be primarily dissipative (appearing in the amplitude quadrature) rather than dispersive (appearing in the phase quadrature).

The relative magnitude of the $b_1$ and $b_2$ responses, which in this operating point reflects the relative dissipative and dispersive coupling, is influenced by the derivative of the reflected light intensity evaluated at the probe detuning frequency. For a cavity whose internal fields are only weakly modified by the sensor, the intensity is insensitive to sensor displacement to first order at zero detuning, making the system purely dispersive. 
However, for our case here, the reflected light intensity varies rapidly with $\phi_f$ at the trap site (because the reflected intensity can vary rapidly with sensor position near the trap). So, the response in $b_1$ still well exceeds that of $b_2$ even when $\Delta=0$. We have also determined numerically that the ratio of dissipative to dispersive coupling is equal to unity for a detuning of $\Delta/\kappa\approx \pm 1$.

With these features in mind, we find that a simple analytic model fits the structure of the transfer function from an external force to the reflected light readout:
\begin{multline}
    \frac{\partial\vec{b}}{\partial F_\text{ext}}(\Omega)\approx i\chi_\text{cooled}(\Omega)A_{\text{probe},1}\\
    \times |\sin(nkL_s)|\vec{\beta}(n,\rho_1,\rho_2,\Delta),
\end{multline}
where $\vec{\beta}$ is a scale set by how well the probe beam sidebands propagate within and eventually escape the cavity (and is thus dictated by the reflectivities of each surface and the detuning).

This structure has a rather intuitive form: a conversion from force on the sensor to its displacement ($\chi_\text{cooled}$), linear scaling with the amplitude of the probe beam itself ($A_{\text{probe},1}$), and a general scale set by optical properties ($\vec{\beta}$). The factor of $i$ shows that the reflected field response picks up an additional $\pi/2$ phase shift relative to the mechanical response of the sensor. However, the conversion from sensor displacement to the readout signal is slightly less intuitive. Naively, this conversion would scale with the optomechanical coupling $g_0$, which we have found scales with $\sin^2(nkL_s)$ when the cavity is locked (see \cite{DielectricLetter}) -- this scaling also matches the power reflectivity of the full dielectric slab \cite{jayich2008dispersive}. However, here we are concerned purely with the AC field amplitude, rather than the differential AC power, which leads to the single power of $\sin(nkL_s)$ instead. 

The probe DC field $\vec B$ as a function of $L_s$ (with the sensor placed at its trap site and the cavity locked) is composed of a large constant value plus an additional term $\propto|\sin(nkL_s)|$. The leading constant value emerges from the portion of the intracavity field which does not interact with the weakly reflective sensor surfaces, instead just seeing the complete sensor as a slight phase gain. The form of $\vec{B}$ results in the total observable transfer function (i.e., the oscillatory differential reflected power) still possessing a term $\propto g_0$; this term is just subdominant relative to the leading $\propto\sqrt{g_0}$ contribution.

\subsubsection{Response to trap laser fluctuations}

In the middle panels of Fig. \ref{fig: example transfer functions}, we plot the transfer functions from fluctuations in the trap field quadrature components to the probe readout channel, using the same parameters -- only inputs $a_1$ and $a_2$ are shown because $p_1$ and $p_2$ generate qualitatively similar results. Even though fluctuations in the trap laser field do not directly couple into the readout signal (as they are sidebands of the trap carrier, rather than the probe), they still can produce differential forces on the sensor which propagate to the readout via the probe field. This observation directly explains the similarity of these transfer functions to those with $\delta F_\text{ext}$ as the input.

The conversion from field sidebands entering the cavity into forces on the sensor due to sidebands merits further exploration. In the limit $\Omega\rightarrow0$, the force from sidebands entering through the input mirror go to zero. In this limit, both the DC quadrature vector $\vec A$ and the sideband vector $\vec a$ experience identical rotation and amplification as they propagate to every other port in the cavity, regardless of $\phi_f$. Furthermore, one can always redefine the temporal  origin so that $A_1\sim\sqrt{P_\text{trap}}$ and $A_2=0$; the fact that $\vec A\cdot\begin{pmatrix} 0 \\ a_2 \end{pmatrix}=0$ ensures that a pure oscillation in $a_2$ provides zero differential power incident on the sensor in the low-frequency limit, and thus $\delta F_\text{trap}=0$. 
Similarly, a field perturbation aligned with the DC input vector (i.e., $a_1$) just leads to a global rescaling of the power at each port in the small $\Omega$ limit. The equilibrium trap location lies where the total optical force is balanced and thus is unaffected by a global power rescaling -- no net force is generated.

Unlike the fluctuations in the trap field entering the input mirror, there is no guarantee that trap laser sidebands $p_1$ and $p_2$ produce no force on the sensor in the $\Omega\rightarrow 0$. The cavity is not necessarily symmetric about the center of the sensor at its trap site (i.e., $[\phi_{e,\text{lock}}-\phi_{f,\text{lock}}-nkL_s]~\text{mod}~2\pi\not=\phi_{f,\text{lock}}$), so the rotation and scaling of $\vec p$ as it propagates to a given port in the cavity ($\vec e$, $\vec f$, etc.) may be different than that of $\vec a$. 
Thus, low-frequency fields entering at $\vec p$ can still produce differential radiation pressure.

However, once $\Omega$ reaches the scales displayed in Fig. \ref{fig: example transfer functions}, the additional rotation angles accumulated by the sideband fields over free space elements leads to non-offsetting differential powers, allowing both $a_1$ and $a_2$ to produce forces which are comparable to those produced by $p_1$ and $p_2$.
From the hierarchy in the response functions, we conclude that fluctuations in the trap beam $a_2$ in the plotted frequency range produce larger forces on the sensor than those in $a_1$. When $f_\text{trap}$ decreases, the hierarchy shifts in favor of $a_1$ --- the additional sideband rotation angle decreases so that sidebands generated from $a_1$ align more closely with the DC fields at the membrane surfaces.
These complexities in the forces produced by external field fluctuations may cause quantum noise in the optical fields to contribute more heavily to the limiting noise sources than perhaps previously expected \cite{LSD2012}; careful attention to this contribution is warranted in future work which estimates the sensitivity of various detector configurations.

\subsubsection{Response to probe laser fluctuations}
In the bottom panels of Fig. \ref{fig: example transfer functions}, we consider the transfer functions from fluctuations in the probe laser quadrature components to the readout port. 
Unlike fluctuations in the trap beam fields, these perturbations can propagate directly to the readout. 
Therefore, these transfer functions contain a direct component that dominates far from the cooled resonance and an indirect component which emerges from the force of the probe on the sensor; in the vicinity of the cooled resonance, these two contributions interfere with one another. 

The direct components for $a_1\rightarrow b_1$ and $a_1\rightarrow b_2$ match those of $a_2\rightarrow b_2$ and $a_2\rightarrow b_1$ (up to a minus sign), respectively. This is generically true once the frequency is far enough away from the resonant frequencies that the sensor motion is negligible. The direct component of the response in $\vec b$ is just a combination of rescalings and rotations of $\vec a$, dictated by how the fields propagate in the cavity, as opposed to acquiring dependencies on the DC field quadrature angles through the dot products contained Eq. \eqref{eq: AC force}. Thus, the quadrature components of the input and output can be converted into one another by a global $\pi/2$ rotation\footnote{This is equivalent to a global shift in the time axis by $\pi/2\omega_0$.}.

Near $f_\text{cooled}$, however, the motion of the dielectric sensor becomes important. From the structures of the transfer functions, we can see that the $a_1\rightarrow b_1$ and $a_1\rightarrow b_2$ functions are dominated by the sensor response in this region, while the $a_2\rightarrow b_1$ and $a_2\rightarrow b_2$ functions only show a small deviation from their constant direct component values. 
The hierarchy in the indirect component's effect in the readout quadrature suggests that relative to the zero detuned case, the hierarchy has flipped --- the force produced by $a_1$ is much larger than that produced by an equivalent $a_2$. 
At $\Delta=-0.5\kappa$, the intracavity probe intensity is highly sensitive to the sensor position. Consequently, incident sideband fields in the amplitude quadrature $a_1$, which perturb the intracavity intensity directly, produce relatively large radiation-pressure forces on the sensor. In contrast, the phase quadrature $a_2$ produces comparatively weak intensity fluctuations and therefore couples less strongly to the force. Significant destructive interference between the direct and indirect components is possible, as evidenced by the sharp localized drop in magnitude for both the pink and gray traces, which are characterized by larger fluctuation-induced forces. 

Finally, for a given laser quadrature ($a_1$ or $a_2$), it is evident that the indirect component (i.e., the deviation from a fixed magnitude) is stronger in the transfer function to $b_1$ than $b_2$. This hierarchy originates from our earlier observation that for this set of model parameters, a force on the sensor propagates primarily to $b_1$.

\subsubsection{Response to cavity mirror displacements}
\begin{figure}
    \centering
    \includegraphics[width=\linewidth]{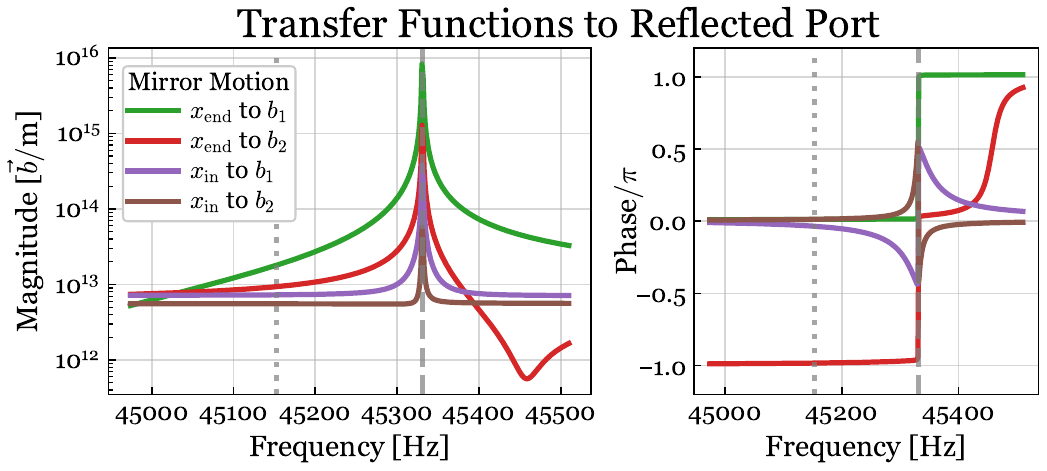}
    \caption{The transfer functions from cavity mirror displacements to the readout channel of the probe beam. Once again, we set $L_s=0.25\lambda/n$ and $\Delta=-0.5\kappa$ and place the sensor at the locked cavity equilibrium. Gray lines are as in Fig. \ref{fig: example transfer functions}.}
    \label{fig: example mirror disp transfer functions}
\end{figure}

The last set of transfer functions we study here are those from displacements of the cavity mirrors to the readout channel, as shown in Fig. \ref{fig: example mirror disp transfer functions}. 
Much like the probe beam fluctuation transfer functions, these response functions are dominated by the indirect component which couples through a force on the sensor near the resonant frequency and, outside that vicinity, by the direct component which does not noticeably vary over mechanical frequency scales. 
In this case, however, the indirect component is dominated by forces from the trap beam, while the direct component still appears only through probe beam sidebands generated by the moving mirrors.

By solving the field equations in an empty cavity, one can show that when the mirrors oscillate at frequencies $\ll\kappa$, the location of the trap site follows the displacement of the end mirror but is insensitive to the displacement of the input mirror \cite{AndrewsGRNote}. 
Therefore, one may expect that near the resonant frequency, the end mirror displacement transfer function should be well approximated by $-m\Omega_{\text{cooled}}^2$ times the external force transfer function, and the input mirror displacement transfer function should be the same, with an additional large suppression factor. 

\begin{figure}[t!]
    \centering
    \includegraphics[width=0.99\linewidth]{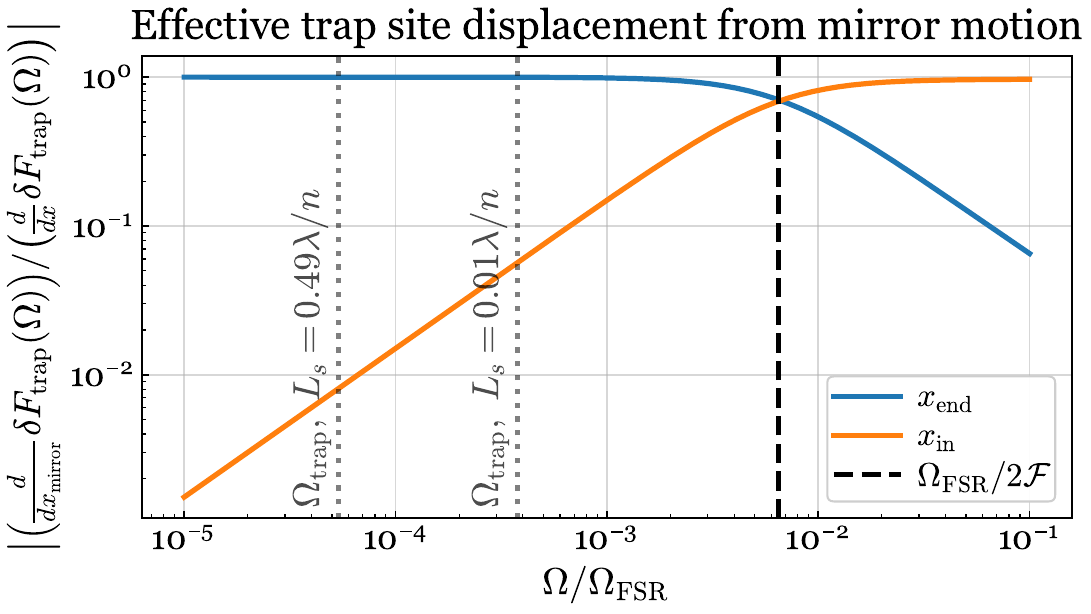}
    \caption{The frequency-dependent ratio between the differential trap laser force due to a small displacement of the cavity mirrors and that due to a displacement of the sensor, which describes the effective displacement of the trap site due to mirror motion.
    The results are \textit{independent of $L_s$} when the cavity is locked and furthermore match those of Fig. 6 of \cite{AndrewsGRNote}, which assumes a small, perturbative sensor.
    For reference, we also note two values of $\Omega_\text{trap}$ to indicate how much input mirror displacement noise suppression may be achieved.}
    \label{fig: force coeff mirror displacement ratio}
\end{figure}

To test the limits of this intuition, we generalize this result to include the effect of the sensor by computing the force coefficients $\frac{\partial}{\partial x_\text{end}}\delta F_\text{trap}(\Omega)$ and $\frac{\partial}{\partial x_\text{in}}\delta F_\text{trap}(\Omega)$ and normalizing by the coefficient\footnote{This term does not appear in \eqref{eq: trap beam differential force} since it would lead to double counting with $\chi_\text{mech}$, but can still be calculated in the same manner as the other coefficients.} $\frac{\partial}{\partial x}\delta F_\text{trap}(\Omega)$. This ratio describes the effective trap site displacement due to the motion of each mirror.
We carry out this computation in Fig. \ref{fig: force coeff mirror displacement ratio}, where we normalize the frequency by the cavity free spectral range.

Comparing the normalized coefficients to the empty cavity calculation in Fig. 6 of \cite{AndrewsGRNote}, we find striking agreement between the two results. In the $\Omega\rightarrow0$ limit, the normalized end mirror force coefficient has a magnitude of one, while its input mirror counterpart approaches zero. 
Furthermore, the input mirror force coefficient ratio surpasses its end mirror counterpart at $\Omega_\text{FSR}/2\mathcal{F}\approx\kappa$, where $\mathcal{F}$ is the cavity finesse, approaching a value very near one. The origin of this behavior in the empty cavity calculation is explained in depth in \cite{AndrewsGRNote}. 

Below $\Omega\approx\kappa$, the leading behavior of the end mirror coefficient ratio is a real constant, while the input mirror counterpart is dominated by an imaginary term scaling with $\Omega$. Above this threshold, the input mirror coefficient ratio is dominated by a real constant while the end mirror coefficient ratio is purely imaginary at leading order, scaling as $\Omega^{-1}$. Thus, at $\Omega\approx\kappa$, not only do the strengths of the two couplings interchange, but their phase relations relative to the trap-displacement response also swap.

Interestingly, despite the absolute optical response depending heavily on the dielectric properties, the coefficient ratios in Fig. \ref{fig: force coeff mirror displacement ratio} are \textit{independent of $L_s$ when the cavity is locked}, meaning that the agreement with the empty cavity calculation of \cite{AndrewsGRNote} holds \textit{even when the sensor sources substantial restructuring of the cavity fields}. We have observed previously that when the cavity is locked and the sensor placed at its stable equilibrium, the power in the subcavities and the cavity linewidths are not only independent of $L_s$, but also match their values for an empty, on-resonance cavity \cite{DielectricLetter}. It is plausible that the striking behavior in the force coefficient ratios would be related to the same physics which generates this $L_s$-independence. This finding strengthens the notion that even though the combined cavity-sensor system displays many departures from standard perturbative expectations, once the cavity is locked, certain aspects emerge where the behavior of the system reunites with the predictions of perturbation theory.

In summary, we have verified that in the vicinity of $\Omega_\text{cooled}$ (i.e., in the region where the indirect component of the full transfer function is dominant), the end mirror transfer function is related to the external force transfer function by a simple conversion factor of $-m\Omega_\text{cooled}^2$, a direct result of the $L_s$-independence in the effective trap site displacement. The input mirror response can be fit with a very similar function, up to a factor determined by the ratio between the two curves plotted in Fig. \ref{fig: force coeff mirror displacement ratio}.

\subsubsection{Response to a gravitational wave}
\label{sec: gw response}

When a high frequency\footnote{``High frequency'' here means that $\Omega$ exceeds the experimental mechanical resonance frequencies but is still $\ll\omega_0$.} gravitational wave passes through the cavity, the length of the cavity oscillates about its mean size. The optical fields inside the cavity respond to the changing length, displacing the trap equilibrium location relative to the position of the sensor. 
In previous work, we showed analytically that to first order in the GW frequency, the displacement of a trap site relative to a very small sensor levitated at that point, $\delta x_\text{GW}$, is gauge-invariant \cite{AndrewsGRNote}. At lowest order, the displacement due to a plus-polarized GW of strain $h(t)=h_0\cos\Omega t$, incident from overhead, is
\begin{equation}
    \delta x_\text{GW}(t)=\frac{h_0}{2}(L_c-L_t)\cos\Omega t.
    \label{eq: delta x antinode sensor left}
\end{equation} 
Considering a coordinate system centered on the equilibrium sensor location and working in the locally-Lorentz gauge, one can show that a GW effectively displaces the end mirror by $h(t)(L_c-L_t)/2$ and the input mirror by $-h(t)L_t/2$. Furthermore, at lowest order in $\Omega$, the trap site follows the end mirror displacement and is insensitive to the input mirror's motion, as seen in Fig. \ref{fig: force coeff mirror displacement ratio}. So for our purposes, where we have left the small particle regime, we can still capture the response of this experiment to a gravitational wave with frequency $\Omega\ll\kappa$ (i.e., generate a perturbation consistent with Eq. \eqref{eq: delta x antinode sensor left}) not by a direct shift of the standing wave pattern, but rather via mirror displacements. In other words, when $\Omega\ll\kappa$ (a valid approximation for established detector designs \cite{LSD2012, LSD2020}), we have
\begin{equation}
    \frac{\partial\vec{b}}{\partial h}=\frac{L_c-L_t}{2}\frac{\partial \vec{b}}{\partial x_\text{end}}-\frac{L_t}{2}\frac{\partial \vec{b}}{\partial x_\text{in}}.
\end{equation}
Note that we have also assumed a plus-polarized wave incident from overhead --- general wave incidence and polarization angles can be incorporated through an additional attenuation factor derived by projecting the GW metric tensor along the detector arms.

Once $\Omega\gtrsim\kappa$, the zeroth-order approximation to the GW-induced displacement no longer holds, and future iterations of this class of experiment may push into this regime \cite{LSD2020}. To properly model a gravitational wave for such configurations, its effects must be encoded as deviations in free space propagation phases, rather than pure mirror displacements. These free space propagation phase deviations can be computed straightforwardly in the manner of \cite{AndrewsGRNote, RakhmanovLLGauge}, but we leave the implementation of this subtlety for future work.

\section{Conclusion}
\label{sec: conclusion}

In this paper, we developed an extension to the vacuum two-photon formalism of \cite{CorbittFramework}, seamlessly incorporating planar dielectric sensors and optical trapping physics. We developed a framework for building dielectrics into the field propagation relations, optical forces, and computations of self-consistent experimental response functions which simultaneously capture the restructuring of the internal cavity fields due to the sensor motion \textit{and} the sensor motion due to the forces from the modified fields. Together, these satisfy the requirements laid out in the Introduction, from the field solution at an arbitrary sensor position to the response to laser fluctuations, mirror displacements, external forces and gravitational waves.

In the accompanying letter \cite{DielectricLetter}, we used this framework to highlight multiple instances of nonperturbative optomechanical physics generated by moderately reflective dielectrics. Here, we also conducted a detailed investigation of the response functions which describe the propagation of external perturbations, both optical and mechanical, to an example readout channel, along with the cooling of the sensor by the trap and probe beams. Notably, once the cavity is locked to maximal transmitted power, the transfer functions from mirror displacements to the readout are related to the external force transfer function by conversion factors which do not depend on the sensor thickness and which match the empty cavity result of \cite{AndrewsGRNote}. This is a result of the same thickness independence of the subcavity powers and cavity linewidths in the locked cavity presented in \cite{DielectricLetter}.

Our treatment assumes plane wave propagation between lossless surfaces of infinite extent, so it cannot capture the off-axis optical trapping behavior of the levitated sensors considered here, nor can it capture the on-axis trapping physics for sensors with thickness equal to an integer multiple of $\lambda/2n$, which are transparent in this model. We also treat a gravitational wave as an effective displacement of the mirrors, which is valid for $\Omega\ll\kappa$; higher frequencies require encoding the wave in the free space propagation phases.

A natural next step is to generalize this framework to more complete experimental models. Accounting for multilayer dielectric sensors or more complicated optical layouts, such as a Fabry-P\'{e}rot Michelson interferometer, only requires additional optical field quadrature vectors and mechanical degrees of freedom. Extending the mathematical framework to account for the full three-dimensional beam profile as it propagates through the optical system represents a greater, but not insurmountable, mathematical challenge. Additionally, the framework admits a straightforward adjustment to replace the levitated dielectrics with stiffly mounted membranes: the mechanical susceptibility and equilibrium location should be described by the mechanical support rather than the local gradient of the trap beam optical force. Notably, this change allows for completely generic combinations of $\phi_f$ and $\phi_e$ for a mount stiff enough that the sensor stays at its mounting position, opening regions of the parameter space which are otherwise inaccessible under the requirement that the sensor lie at a stable equilibrium set exclusively by optical forces. The intermediate stiffness regime is discussed in the Supplemental Material of \cite{DielectricLetter}.

The ultimate goal of this campaign is a set of estimates for the experimental GW strain sensitivity. The formalism developed in this work is the prerequisite for that calculation: it supplies the transfer functions needed to project each noise source onto the readout and normalize to strain, and without it a noise budget for a sensor in this regime cannot be computed self-consistently. The complete noise budget calculation will then provide a means for systematically exploring the detector design parameter space and identifying optimal target configurations.

\begin{acknowledgments}
A.L. is grateful for support from the Fannie and John Hertz Foundation through a Hertz Fellowship. N.A. is supported by the UC Noyce Initiative and the Department of Energy award number DE-SC0009999. We also thank Rana Adhikari, Dan Brown, Yanbei Chen, Su Direkci, Vincent Dumont, Shafaq Elahi, James Gardner, Andrew Geraci, Shane Larson, Aaron Markowitz, Kevin Kuns, Jack Sankey, Vivishek Sudhir, Christopher Wipf, and Dalziel Wilson for fruitful discussions throughout this work.
\end{acknowledgments}

\bibliography{main}

@article{RakhmanovLLGauge,
    author = "Rakhmanov, Malik",
    title = "{Response of test masses to gravitational waves in the local Lorentz gauge}",
    eprint = "gr-qc/0406009",
    archivePrefix = "arXiv",
    reportNumber = "LIGO-P030056",
    doi = "10.1103/PhysRevD.71.084003",
    journal = "Phys. Rev. D",
    volume = "71",
    pages = "084003",
    year = "2005"
}

@phdthesis{AggarwalThesis,
    author = "Aggarwal, Nancy",
    title = "{A room temperature optomechanical squeezer}",
    eprint = "2006.14323",
    archivePrefix = "arXiv",
    primaryClass = "quant-ph",
    doi = "10.1038/s41567-020-0877-x",
    school = "MIT",
    month = "2",
    year = "2019"
}

@article{LSD2020,
    author = "Aggarwal, Nancy and Winstone, George P. and Teo, Mae and Baryakhtar, Masha and Larson, Shane L. and Kalogera, Vicky and Geraci, Andrew A.",
    title = "{Searching for New Physics with a Levitated-Sensor-Based Gravitational-Wave Detector}",
    eprint = "2010.13157",
    archivePrefix = "arXiv",
    primaryClass = "gr-qc",
    doi = "10.1103/PhysRevLett.128.111101",
    journal = "Phys. Rev. Lett.",
    volume = "128",
    number = "11",
    pages = "111101",
    year = "2022"
}

@article{LSD2022,
    author = "Winstone, George and others",
    collaboration = "LSD",
    title = "{Optical Trapping of High-Aspect-Ratio NaYF Hexagonal Prisms for kHz-MHz Gravitational Wave Detectors}",
    eprint = "2204.10843",
    archivePrefix = "arXiv",
    primaryClass = "physics.optics",
    doi = "10.1103/PhysRevLett.129.053604",
    journal = "Phys. Rev. Lett.",
    volume = "129",
    number = "5",
    pages = "053604",
    year = "2022"
}

@ARTICLE{Byrnes2016,
       author = {{Byrnes}, Steven J.},
        title = "{Multilayer optical calculations}",
      journal = {arXiv e-prints},
         year = 2016,
        month = mar,
          eid = {arXiv:1603.02720},
        pages = {arXiv:1603.02720},
          doi = {10.48550/arXiv.1603.02720},
archivePrefix = {arXiv},
       eprint = {1603.02720},
 primaryClass = {physics.comp-ph},
       adsurl = {https://ui.adsabs.harvard.edu/abs/2016arXiv160302720B}
}

@article{CorbittFramework,
    author = "Corbitt, Thomas and Chen, Yanbei and Mavalvala, Nergis",
    title = "{Mathematical framework for simulation of quantum fields in complex interferometers using the two-photon formalism}",
    eprint = "quant-ph/0502088",
    archivePrefix = "arXiv",
    reportNumber = "LIGO-P030070-00-R, AEI-2005-006",
    doi = "10.1103/PhysRevA.72.013818",
    journal = "Phys. Rev. A",
    volume = "72",
    pages = "013818",
    year = "2005"
}

@article{LSD2012,
    author = "Arvanitaki, Asimina and Geraci, Andrew A.",
    title = "{Detecting high-frequency gravitational waves with optically-levitated sensors}",
    eprint = "1207.5320",
    archivePrefix = "arXiv",
    primaryClass = "gr-qc",
    doi = "10.1103/PhysRevLett.110.071105",
    journal = "Phys. Rev. Lett.",
    volume = "110",
    number = "7",
    pages = "071105",
    year = "2013"
}

@article{ChangThinDiscs,
       author = {{Chang}, D.~E. and {Ni}, K. -K. and {Painter}, O. and {Kimble}, H.~J.},
        title = "{Ultrahigh-Q mechanical oscillators through optical trapping}",
      journal = {arXiv e-prints},
         year = 2010,
        month = dec,
          eid = {arXiv:1101.0146},
        pages = {arXiv:1101.0146},
          doi = {10.48550/arXiv.1101.0146},
archivePrefix = {arXiv},
       eprint = {1101.0146},
 primaryClass = {quant-ph},
       adsurl = {https://ui.adsabs.harvard.edu/abs/2011arXiv1101.0146C}
}

@ARTICLE{ChangSpheres,
       author = {{Chang}, D.~E. and {Regal}, C.~A. and {Papp}, S.~B. and {Wilson}, D.~J. and {Ye}, J. and {Painter}, O. and {Kimble}, H.~J. and {Zoller}, P.},
        title = "{Cavity opto-mechanics using an optically levitated nanosphere}",
      journal = {Proceedings of the National Academy of Science},
         year = 2010,
        month = jan,
       volume = {107},
       number = {3},
        pages = {1005-1010},
          doi = {10.1073/pnas.0912969107},
archivePrefix = {arXiv},
       eprint = {0909.1548},
 primaryClass = {quant-ph},
       adsurl = {https://ui.adsabs.harvard.edu/abs/2010PNAS..107.1005C}
}

@phdthesis{DumontThesis,
    author = "Dumont, Vincent",
    title = "{Quantum noise and mechanical membranes in microcavities}",
    year = "2022"
}

@article{CavesAndSchumaker1,
    author = "Caves, Carlton M. and Schumaker, Bonny L.",
    title = "{New formalism for two-photon quantum optics. 1. Quadrature phases and squeezed states}",
    doi = "10.1103/PhysRevA.31.3068",
    journal = "Phys. Rev. A",
    volume = "31",
    pages = "3068--3092",
    year = "1985"
}

@article{CavesAndSchumaker2,
    author = "Schumaker, Bonny L. and Caves, Carlton M.",
    title = "{New formalism for two-photon quantum optics. 2. Mathematical foundation and compact notation}",
    doi = "10.1103/PhysRevA.31.3093",
    journal = "Phys. Rev. A",
    volume = "31",
    pages = "3093--3111",
    year = "1985"
}

@book{Bak2010,
  title     = "Complex analysis",
  author    = "Bak, Joseph and Newman, Donald J",
  publisher = "Springer",
  series    = "Undergraduate texts in mathematics",
  edition   =  3,
  month     =  aug,
  year      =  2010,
  address   = "New York, NY"
}

@article{CavityOptomechanics,
    author = "Aspelmeyer, Markus and Kippenberg, Tobias J. and Marquardt, Florian",
    title = "{Cavity Optomechanics}",
    eprint = "1303.0733",
    archivePrefix = "arXiv",
    primaryClass = "cond-mat.mes-hall",
    doi = "10.1103/RevModPhys.86.1391",
    journal = "Rev. Mod. Phys.",
    volume = "86",
    pages = "1391",
    year = "2014"
}

@article{KieselCooling,
    author = "Kiesel, Nikolai and Blaser, Florian and Delic, Uros and Grass, David and Kaltenbaek, Rainer and Aspelmeyer, Markus",
    title = "{Cavity cooling of an optically levitated nanoparticle}",
    eprint = "1304.6679",
    archivePrefix = "arXiv",
    primaryClass = "quant-ph",
    doi = "10.1073/pnas.1309167110",
    journal = "Proc. Nat. Acad. Sci.",
    volume = "11",
    pages = "14180--14185",
    year = "2013"
}

@article{FonsecaCooling,
    author = "Fonseca, P. Z. G. and Aranas, E. B. and Millen, J. and Monteiro, T. S. and Barker, P. F.",
    title = "{Nonlinear dynamics and millikelvin cavity-cooling of levitated nanoparticles}",
    eprint = "1511.08482",
    archivePrefix = "arXiv",
    primaryClass = "quant-ph",
    doi = "10.1103/PhysRevLett.117.173602",
    journal = "Phys. Rev. Lett.",
    volume = "117",
    pages = "173602",
    year = "2016"
}

@article{Delic:2019xqd,
    author = "Deli{\'c}, Uro{\v{s}} and Reisenbauer, Manuel and Dare, Kahan and Grass, David and Vuleti{\'c}, Vladan and Kiesel, Nikolai and Aspelmeyer, Markus",
    title = "{Cooling of a levitated nanoparticle to the motional quantum ground state}",
    eprint = "1911.04406",
    archivePrefix = "arXiv",
    primaryClass = "quant-ph",
    doi = "10.1126/science.aba3993",
    journal = "Science",
    volume = "367",
    number = "6480",
    pages = "892--895",
    year = "2020"
}

@article{Delic:2019gri,
    author = "Deli{\'c}, Uro{\v{s}} and Reisenbauer, Manuel and Grass, David and Kiesel, Nikolai and Vuleti{\'c}, Vladan and Aspelmeyer, Markus",
    title = "{Cavity cooling of a levitated nanosphere by coherent scattering}",
    eprint = "1812.09358",
    archivePrefix = "arXiv",
    primaryClass = "quant-ph",
    doi = "10.1103/PhysRevLett.122.123602",
    journal = "Phys. Rev. Lett.",
    volume = "122",
    pages = "123602",
    year = "2019"
}

@article{Pontin:2022ywq,
    author = "Pontin, Antonio and Fu, Hayden and Toro{\v{s}}, Marko and Monteiro, Tania S. and Barker, Peter F.",
    title = "{Simultaneous cavity cooling of all six degrees of freedom of a levitated nanoparticle}",
    eprint = "2205.10193",
    archivePrefix = "arXiv",
    primaryClass = "quant-ph",
    doi = "10.1038/s41567-023-02006-6",
    journal = "Nature Phys.",
    volume = "19",
    number = "7",
    pages = "1003--1008",
    year = "2023"
}

@article{Windey:2019yoi,
    author = "Windey, Dominik and Gonzalez-Ballestero, Carlos and Maurer, Patrick and Novotny, Lukas and Romero-Isart, Oriol and Reimann, Ren{\'e}",
    title = "{Cavity-Based 3D Cooling of a Levitated Nanoparticle via Coherent Scattering}",
    eprint = "1812.09176",
    archivePrefix = "arXiv",
    primaryClass = "quant-ph",
    doi = "10.1103/PhysRevLett.122.123601",
    journal = "Phys. Rev. Lett.",
    volume = "122",
    pages = "123601",
    year = "2019"
}

@ARTICLE{AndrewsGRNote,
       author = {{Laeuger}, Andrew and {Elahi}, Shafaq Gulzar and {Klomp}, Shelby and {Larsen}, Jackson and {Sprague}, Jacob and {Wang}, Zhiyuan and {Winstone}, George and {Wroblewski}, Maddox and {Larson}, Shane L. and {Geraci}, Andrew A. and {Aggarwal}, Nancy},
        title = "{Gravitational wave signal and noise response of an optically levitated sensor in a Fabry-P{\'e}rot cavity}",
      journal = {\prd},
         year = 2026,
        month = sep,
       volume = {114},
       number = {6},
          eid = {064071},
        pages = {064071},
          doi = {10.1103/mhz8-h774},
archivePrefix = {arXiv},
       eprint = {2604.07302},
 primaryClass = {gr-qc},
}

@article{Dumont:2019ziy,
    author = "Dumont, Vincent and Bernard, Simon and Reinhardt, Christoph and Kato, Alex and Ruf, Maximilian and Sankey, Jack C.",
    journal = {Opt. Express},
    number = {18},
    pages = {25731--25748},
    publisher = {Optica Publishing Group},
    volume = {27},
    month = {Sep},
    year = {2019},
    title = "{Flexure-Tuned Membrane-at-the-Edge Optomechanical System}",
    url = {https://opg.optica.org/oe/abstract.cfm?URI=oe-27-18-25731},
    eprint = "1905.04594",
    archivePrefix = "arXiv",
    primaryClass = "quant-ph",
    doi = "10.1364/OE.27.025731",
}

@article{Thompson:2007hno,
    author = "Thompson, J. D. and Zwickl, B. M. and Jayich, A. M. and Marquardt, Florian and Girvin, S. M. and Harris, J. G. E.",
    title = "{Strong dispersive coupling of a high-finesse cavity to a micromechanical membrane}",
    eprint = "0707.1724",
    archivePrefix = "arXiv",
    primaryClass = "quant-ph",
    doi = "10.1038/nature06715",
    journal = "Nature",
    volume = "452",
    number = "7183",
    pages = "72--75",
    year = "2008"
}

@article{xu2014squeezing,
  title={Squeezing in a coupled two-mode optomechanical system for force sensing below the standard quantum limit},
  author={Xu, Xunnong and Taylor, Jacob M},
  journal={Physical Review A},
  volume={90},
  number={4},
  pages={043848},
  year={2014},
  publisher={APS},
  arxivPrefix = "arXiv",
  eprint = "1303.7469",
  doi = "10.1103/PhysRevA.90.043848"
}

@article{xuereb2012dynamical,
  title={Dynamical scattering models in optomechanics: going beyond the ‘coupled cavities’ model},
  author={Xuereb, Andr{\'e} and Domokos, Peter},
  journal={New Journal of Physics},
  volume={14},
  number={9},
  pages={095027},
  year={2012},
  publisher={IOP Publishing},
  doi = "10.1088/1367-2630/14/9/095027",
  
}

@article{xuereb2010optomechanical,
  title={Optomechanical cooling with generalized interferometers},
  author={Xuereb, Andr{\'e} and Freegarde, Tim and Horak, Peter and Domokos, Peter},
  journal={Physical review letters},
  volume={105},
  number={1},
  pages={013602},
  year={2010},
  publisher={APS},
  doi = "10.1103/PhysRevLett.105.013602",
  arxivPrefix = "arXiv",
  eprint = "1002.0463"

}

@article{stambaugh2015membrane,
  title={From membrane-in-the-middle to mirror-in-the-middle with a high-reflectivity sub-wavelength grating},
  author={Stambaugh, Corey and Xu, Haitan and Kemiktarak, Utku and Taylor, Jacob and Lawall, John},
  journal={Annalen der Physik},
  volume={527},
  number={1-2},
  pages={81--88},
  year={2015},
  publisher={Wiley Online Library},
  eprint = "1407.1709",
  archivePrefix = "arXiv",
  primaryClass = "physics.optics",
}

@article{jayich2008dispersive,
    author = "Harris, J. G E. and Marquardt, F. and Clerk, A. A. and Girvin, S. M. and Thompson, J. D. and Yang, C. and Zwickl, B. M. and Sankey, J. C. and Jayich, A. M.",
    title = "{Dispersive optomechanics: a membrane inside a cavity}",
    eprint = "0805.3723",
    archivePrefix = "arXiv",
    primaryClass = "quant-ph",
    doi = "10.1088/1367-2630/10/9/095008",
    journal = "New J. Phys.",
    volume = "10",
    number = "9",
    pages = "095008",
    year = "2008"
}

@article{Ashkin:86,
author = {A. Ashkin and J. M. Dziedzic and J. E. Bjorkholm and Steven Chu},
journal = {Opt. Lett.},
number = {5},
pages = {288--290},
publisher = {Optica Publishing Group},
title = {Observation of a single-beam gradient force optical trap for dielectric particles},
volume = {11},
month = {May},
year = {1986},
url = {https://opg.optica.org/ol/abstract.cfm?URI=ol-11-5-288},
doi = {10.1364/OL.11.000288},
}

@article{Ashkin:1978zsu,
    author = "Ashkin, A.",
    title = "{Trapping of Atoms by Resonance Radiation Pressure}",
    doi = "10.1103/PhysRevLett.40.729",
    journal = "Phys. Rev. Lett.",
    volume = "40",
    number = "12",
    pages = "729--732",
    year = "1978"
}

@article{Gordon,
  title = {Radiation Forces and Momenta in Dielectric Media},
  author = {Gordon, James P.},
  journal = {Phys. Rev. A},
  volume = {8},
  issue = {1},
  pages = {14--21},
  numpages = {0},
  year = {1973},
  month = {Jul},
  publisher = {American Physical Society},
  doi = {10.1103/PhysRevA.8.14},
  url = {https://link.aps.org/doi/10.1103/PhysRevA.8.14}
}

@ARTICLE{GordonplusAshkin,
       author = {{Gordon}, J.~P. and {Ashkin}, A.},
        title = "{Motion of atoms in a radiation trap}",
      journal = {\pra},
         year = 1980,
        month = may,
       volume = {21},
       number = {5},
        pages = {1606-1617},
          doi = {10.1103/PhysRevA.21.1606},
}

@ARTICLE{PDHlocking,
       author = {{Drever}, R.~W.~P. and {Hall}, J.~L. and {Kowalski}, F.~V. and {Hough}, J. and {Ford}, G.~M. and {Munley}, A.~J. and {Ward}, H.},
        title = "{Laser phase and frequency stabilization using an optical resonator}",
      journal = {Applied Physics B: Lasers and Optics},
         year = 1983,
        month = jun,
       volume = {31},
       number = {2},
        pages = {97-105},
          doi = {10.1007/BF00702605},
}

@article{Rousso:2026duo,
    author = "Rousso, David and Kunze, Moritz Bjoern Kristiansson and Reinhardt, Christoph",
    title = "{Optomechanical platform for high-frequency gravitational wave and vector dark matter detection}",
    eprint = "2601.02576",
    archivePrefix = "arXiv",
    primaryClass = "astro-ph.IM",
    month = "1",
    year = "2026",
    journal = {}
}

@article{Kippenburg2007,
       author = {{Kippenberg}, Tobias J. and {Vahala}, Kerry J.},
        title = "{Cavity Opto-Mechanics}",
      journal = {Optics Express},
         year = 2007,
        month = dec,
       volume = {15},
       number = {25},
        pages = {17172},
          doi = {10.1364/OE.15.017172},
archivePrefix = {arXiv},
       eprint = {0712.1618},
 primaryClass = {physics.optics},
}

@article{Metcalfe2014,
    author = {Metcalfe, Michael},
    title = {Applications of cavity optomechanics},
    journal = {Applied Physics Reviews},
    volume = {1},
    number = {3},
    pages = {031105},
    year = {2014},
    month = {09},
    issn = {1931-9401},
    doi = {10.1063/1.4896029},
    url = {https://doi.org/10.1063/1.4896029},
}

@article{Li2021,
author = {Li, Bei-Bei and Ou, Lingfeng and Lei, Yuechen and Liu, Yong-Chun},
title = {Cavity optomechanical sensing},
journal = {Nanophotonics},
volume = {10},
number = {11},
pages = {2799-2832},
doi = {https://doi.org/10.1515/nanoph-2021-0256},
year = {2021}
}

@ARTICLE{Millen2020,
       author = {{Millen}, James and {Monteiro}, Tania S. and {Pettit}, Robert and {Vamivakas}, A. Nick},
        title = "{Optomechanics with levitated particles}",
      journal = {Reports on Progress in Physics},
         year = 2020,
        month = feb,
       volume = {83},
       number = {2},
          eid = {026401},
        pages = {026401},
          doi = {10.1088/1361-6633/ab6100},
archivePrefix = {arXiv},
       eprint = {1907.08198},
 primaryClass = {physics.optics},
}

@article{Ranjit:2016dlq,
    author = "Ranjit, Gambhir and Cunningham, Mark and Casey, Kirsten and Geraci, Andrew A.",
    title = "{Zeptonewton force sensing with nanospheres in an optical lattice}",
    eprint = "1603.02122",
    archivePrefix = "arXiv",
    primaryClass = "physics.optics",
    doi = "10.1103/PhysRevA.93.053801",
    journal = "Phys. Rev. A",
    volume = "93",
    number = "5",
    pages = "053801",
    year = "2016"
}

@article{Montoya:2021dqz,
    author = "Montoya, Cris and Alejandro, Eduardo and Eom, William and Grass, Daniel and Clarisse, Nicolas and Witherspoon, Apryl and Geraci, Andrew A.",
    title = "{Scanning force sensing at micrometer distances from a conductive surface with nanospheres in an optical lattice}",
    eprint = "2103.03420",
    archivePrefix = "arXiv",
    primaryClass = "physics.optics",
    doi = "10.1364/AO.457148",
    journal = "Appl. Opt.",
    volume = "61",
    number = "12",
    pages = "3486--3493",
    year = "2022"
}

@article{Hebestreit:2017xur,
    author = "Hebestreit, Erik and Frimmer, Martin and Reimann, Ren{\'e} and Novotny, Lukas",
    title = "{Sensing Static Forces with Free-Falling Nanoparticles}",
    eprint = "1801.01169",
    archivePrefix = "arXiv",
    primaryClass = "physics.ins-det",
    doi = "10.1103/PhysRevLett.121.063602",
    journal = "Phys. Rev. Lett.",
    volume = "121",
    number = "6",
    pages = "063602",
    year = "2018"
}

@article{Monteiro:2020qiz,
    author = "Monteiro, Fernando and Li, Wenqiang and Afek, Gadi and Li, Chang-ling and Mossman, Michael and Moore, David C.",
    title = "{Force and acceleration sensing with optically levitated nanogram masses at microkelvin temperatures}",
    eprint = "2001.10931",
    archivePrefix = "arXiv",
    primaryClass = "physics.optics",
    doi = "10.1103/PhysRevA.101.053835",
    journal = "Phys. Rev. A",
    volume = "101",
    number = "5",
    pages = "053835",
    year = "2020"
}

@article{Kawasaki:2020oyl,
    author = "Kawasaki, Akio and Fieguth, Alexander and Priel, Nadav and Blakemore, Charles P. and Martin, Denzal and Gratta, Giorgio",
    title = "{High sensitivity, levitated microsphere apparatus for short-distance force measurements}",
    eprint = "2004.10973",
    archivePrefix = "arXiv",
    primaryClass = "physics.ins-det",
    doi = "10.1063/5.0011759",
    journal = "Rev. Sci. Instrum.",
    volume = "91",
    number = "8",
    pages = "083201",
    year = "2020"
}

@article{Geraci:2010ft,
    author = "Geraci, Andrew A. and Papp, Scott B. and Kitching, John",
    title = "{Short-range force detection using optically-cooled levitated microspheres}",
    eprint = "1006.0261",
    archivePrefix = "arXiv",
    primaryClass = "hep-ph",
    doi = "10.1103/PhysRevLett.105.101101",
    journal = "Phys. Rev. Lett.",
    volume = "105",
    pages = "101101",
    year = "2010"
}

@article{Rabl:2011nyt,
    author = "Rabl, Peter",
    title = "{Photon Blockade Effect in Optomechanical Systems}",
    eprint = "1102.0278",
    archivePrefix = "arXiv",
    primaryClass = "quant-ph",
    doi = "10.1103/PhysRevLett.107.063601",
    journal = "Phys. Rev. Lett.",
    volume = "107",
    number = "6",
    pages = "063601",
    year = "2011"
}

@ARTICLE{Marquardt2009,
       author = {{Marquardt}, Florian and {Girvin}, S.~M.},
        title = "{Optomechanics (a brief review)}",
      journal = {arXiv e-prints},
         year = 2009,
        month = may,
          eid = {arXiv:0905.0566},
archivePrefix = {arXiv},
       eprint = {0905.0566},
 primaryClass = {cond-mat.mes-hall},
}

@article{Maurer:2022nvj,
    author = "Maurer, Patrick and Gonzalez-Ballestero, Carlos and Romero-Isart, Oriol",
    title = "{Quantum theory of light interaction with a Lorenz-Mie particle: Optical detection and three-dimensional ground-state cooling}",
    eprint = "2212.04838",
    archivePrefix = "arXiv",
    primaryClass = "physics.optics",
    doi = "10.1103/PhysRevA.108.033714",
    journal = "Phys. Rev. A",
    volume = "108",
    number = "3",
    pages = "033714",
    year = "2023",
    note = "[Erratum: Phys.Rev.A 109, 019901 (2024)]"
}

@article{Maurer:2021zgx,
    author = "Maurer, Patrick and Gonzalez-Ballestero, Carlos and Romero-Isart, Oriol",
    title = "{Quantum electrodynamics with a nonmoving dielectric sphere: quantizing Lorenz{\textendash}Mie scattering}",
    eprint = "2106.07975",
    archivePrefix = "arXiv",
    primaryClass = "quant-ph",
    doi = "10.1364/JOSAB.498540",
    journal = "J. Opt. Soc. Am. B",
    volume = "40",
    number = "12",
    pages = "3137--3155",
    year = "2023"
}

@article{Reigue:2022thv,
    author = "Reigue, Antoine and Fogliano, Francesco and Heringlake, Philip and Mercier de L{\'e}pinay, Laure and Besga, Benjamin and Reichel, Jakob and Pigeau, Benjamin and Arcizet, Olivier",
    title = "{Cavity Nano-Optomechanics with Suspended Subwavelength-Sized Nanowires}",
    eprint = "2212.02860",
    archivePrefix = "arXiv",
    primaryClass = "quant-ph",
    doi = "10.1103/PhysRevApplied.20.014025",
    journal = "Phys. Rev. Applied",
    volume = "20",
    number = "1",
    pages = "014025",
    year = "2023"
}

@ARTICLE{Komori2020,
       author = {{Komori}, Kentaro and {Enomoto}, Yutaro and {Ooi}, Ching Pin and {Miyazaki}, Yuki and {Matsumoto}, Nobuyuki and {Sudhir}, Vivishek and {Michimura}, Yuta and {Ando}, Masaki},
        title = "{Attonewton-meter torque sensing with a macroscopic optomechanical torsion pendulum}",
      journal = {\pra},
         year = 2020,
        month = jan,
       volume = {101},
       number = {1},
          eid = {011802},
        pages = {011802},
          doi = {10.1103/PhysRevA.101.011802},
archivePrefix = {arXiv},
       eprint = {1907.13139},
 primaryClass = {quant-ph},
}

@ARTICLE{Gavartin2012,
       author = {{Gavartin}, E. and {Verlot}, P. and {Kippenberg}, T.~J.},
        title = "{A hybrid on-chip optomechanical transducer for ultrasensitive force measurements}",
      journal = {Nature Nanotechnology},
         year = 2012,
        month = aug,
       volume = {7},
       number = {8},
        pages = {509-514},
          doi = {10.1038/nnano.2012.97},
archivePrefix = {arXiv},
       eprint = {1112.0797},
 primaryClass = {physics.optics},
}

@article{Zhou2021,
author = {Feng Zhou and Yiliang Bao and Ramgopal Madugani and David A. Long and Jason J. Gorman and Thomas W. LeBrun},
journal = {Optica},
number = {3},
pages = {350--356},
publisher = {Optica Publishing Group},
title = {Broadband thermomechanically limited sensing with an optomechanical accelerometer},
volume = {8},
month = {Mar},
year = {2021},
url = {https://opg.optica.org/optica/abstract.cfm?URI=optica-8-3-350},
doi = {10.1364/OPTICA.413117},
}

@article{Liang2023,
title = {Yoctonewton force detection based on optically levitated oscillator},
journal = {Fundamental Research},
volume = {3},
number = {1},
pages = {57-62},
year = {2023},
issn = {2667-3258},
doi = {https://doi.org/10.1016/j.fmre.2022.09.021},
url = {https://www.sciencedirect.com/science/article/pii/S2667325822003879},
author = {Tao Liang and Shaochong Zhu and Peitong He and Zhiming Chen and Yingying Wang and Cuihong Li and Zhenhai Fu and Xiaowen Gao and Xinfan Chen and Nan Li and Qi Zhu and Huizhu Hu},
}

@article{Ahn:2019cbu,
    author = "Ahn, Jonghoon and Xu, Zhujing and Bang, Jaehoon and Ju, Peng and Gao, Xingyu and Li, Tongcang",
    title = "{Ultrasensitive torque detection with an optically levitated nanorotor}",
    eprint = "1908.03453",
    archivePrefix = "arXiv",
    primaryClass = "physics.app-ph",
    doi = "10.1038/s41565-019-0605-9",
    journal = "Nature Nanotech.",
    volume = "15",
    pages = "89",
    year = "2020"
}

@article{Ju:2023egl,
    author = "Ju, Peng and Jin, Yuanbin and Shen, Kunhong and Duan, Yao and Xu, Zhujing and Gao, Xingyu and Ni, Xinjie and Li, Tongcang",
    title = "{Near-field GHz rotation and sensing with an optically levitated nanodumbbell}",
    eprint = "2301.10868",
    archivePrefix = "arXiv",
    primaryClass = "quant-ph",
    doi = "10.1021/acs.nanolett.3c02442",
    journal = "Nano Lett.",
    volume = "23",
    pages = "10157",
    year = "2023"
}

@article{Guha2020,
  title = {Force Sensing with an Optomechanical Self-Oscillator},
  author = {Guha, Biswarup and Allain, Pierre Etienne and Lema\^{\i}tre, Aristide and Leo, Giuseppe and Favero, Ivan},
  journal = {Phys. Rev. Appl.},
  volume = {14},
  issue = {2},
  pages = {024079},
  numpages = {11},
  year = {2020},
  month = {Aug},
  publisher = {American Physical Society},
  doi = {10.1103/PhysRevApplied.14.024079},
  url = {https://link.aps.org/doi/10.1103/PhysRevApplied.14.024079}
}

@misc{DielectricLetter,
    author = "Laeuger, Andrew and Aggarwal, Nancy",
    title = "Nonperturbative cavity optomechanics with thin, weakly reflective intracavity dielectrics",
    year={2026}
    }

@misc{ForceLetter,
    author = "Laeuger, Andrew and Chen, Yanbei and Aggarwal, Nancy",
    title = "The optical force on a dielectric slab",
    year={2026}
    }

@misc{DiskTrappingPaper,
    author = "Klomp, Shelby and Elahi, Shafaq Gulzar and Grudichak, Scott and Galla, Chethn and Wang, Zhiyuan and Nguyen, Mark and Winstone, George and Guy, Katarina Boskovic and Derrick, Emma and Wroblewski, Maddox and Sprague, Jacob and Laeuger, Andrew and Larsen, Jackson and Aggarwal, Nancy and Larson, Shane and Geraci, Andrew A.",
    title = "Trapping of nanofabricated {SiO$_2$} high-aspect-ratio discs in an optical lattice",
    year = {2026},
    note = "in prep"
}

@article{Cripe:2017cvb,
    author = "Cripe, Jonathan and Aggarwal, Nancy and Singh, Robinjeet and Lanza, Robert and Libson, Adam and Yap, Min Jet and Cole, Garrett D. and McClelland, David E. and Mavalvala, Nergis and Corbitt, Thomas",
    title = "{Radiation-Pressure-Mediated Control of an Optomechanical Cavity}",
    eprint = "1710.04700",
    archivePrefix = "arXiv",
    primaryClass = "quant-ph",
    doi = "10.1103/PhysRevA.97.013827",
    journal = "Phys. Rev. A",
    volume = "97",
    number = "1",
    pages = "013827",
    year = "2018"
}

@article{cripe2019measurement,
  title={Measurement of quantum back action in the audio band at room temperature},
  author={Cripe, Jonathan and Aggarwal, Nancy and Lanza, Robert and Libson, Adam and Singh, Robinjeet and Heu, Paula and Follman, David and Cole, Garrett D and Mavalvala, Nergis and Corbitt, Thomas},
  journal={Nature},
  volume={568},
  number={7752},
  pages={364--367},
  year={2019},
  publisher={Nature Publishing Group UK London}
}

@article{aggarwal2020room,
  title={Room-temperature optomechanical squeezing},
  author={Aggarwal, Nancy and Cullen, Torrey J and Cripe, Jonathan and Cole, Garrett D and Lanza, Robert and Libson, Adam and Follman, David and Heu, Paula and Corbitt, Thomas and Mavalvala, Nergis},
  journal={Nature Physics},
  volume={16},
  number={7},
  pages={784--788},
  year={2020},
  publisher={Nature Publishing Group UK London}
}

@ARTICLE{Dana2026,
       author = {{Dana}, Andrew M. and {Borden}, Samuel J. and {Burrell}, Nia and {Galla}, Chethn and {Grinin}, Alexey and {Nguyen}, Mark and {Menon}, Kristina and {Elahi}, Shafaq and {Boskovic Guy}, Katarina and {Laeuger}, Andrew and {Weisman}, Evan and {Ranjit}, Gambhir and {Geraci}, Andrew A.},
        title = "{Fiber-based scanning levitated nanoparticle force sensor near a metalized membrane}",
      journal = {arXiv e-prints},
         year = 2026,
        month = sep,
          eid = {arXiv:2609.29829},
        pages = {arXiv:2609.29829},
          doi = {10.48550/arXiv.2609.29829},
archivePrefix = {arXiv},
       eprint = {2609.29829},
 primaryClass = {physics.optics},
}

\appendix

\section{Cavity fields system of equations}
\label{app: full field equations}

In this Appendix, we provide the system of equations which describes the configuration of fields for the experimental model appearing in Fig. \ref{fig:fields diagram complete} and restate the definitions of parameters appearing therein.

The DC fields in the cavity satisfy the following equations:

\begin{subequations}\label{eq:side-by-side-numbered}
\vspace{-16pt}

\noindent
\begin{minipage}[t]{0.48\columnwidth}
\begin{equation}
    \vec B=\rho\vec A+\tau\vec D,~~~~~
\end{equation}
\end{minipage}\hfill
\begin{minipage}[t]{0.48\columnwidth}
\begin{equation}
    \vec C=\tau\vec A-\rho\vec D,~~~
\end{equation}
\end{minipage}

\vspace{-12pt}

\noindent
\begin{minipage}[t]{0.48\columnwidth}
\begin{equation}
    \vec D=\mathbb{R}(\phi_1)\vec F,~~~~~~~~
\end{equation}
\end{minipage}\hfill
\begin{minipage}[t]{0.48\columnwidth}
\begin{equation}
    \vec E=\mathbb{R}(\phi_1)\vec C,~~~~~
\end{equation}
\end{minipage}

\vspace{-12pt}

\noindent
\begin{minipage}[t]{0.48\columnwidth}
\begin{equation}
    \vec F=-\rho_s\vec E+\tau_s^o\vec H,
\end{equation}
\end{minipage}\hfill
\begin{minipage}[t]{0.48\columnwidth}
\begin{equation}
    \vec G=\tau_s^i\vec E+\rho_s\vec H,~
\end{equation}
\end{minipage}

\vspace{-12pt}

\noindent
\begin{minipage}[t]{0.48\columnwidth}
\begin{equation}
    \vec H=\mathbb{R}(\bar{k}L_s)\vec J,~~~~~~~
\end{equation}
\end{minipage}\hfill
\begin{minipage}[t]{0.48\columnwidth}
\begin{equation}
    \vec I=\mathbb{R}(\bar{k}L_s)\vec G,~~
\end{equation}
\end{minipage}

\vspace{-12pt}

\noindent
\begin{minipage}[t]{0.48\columnwidth}
\begin{equation}
     \vec J=\rho_s\vec I+\tau_s^i\vec L,~~~~
\end{equation}
\end{minipage}\hfill
\begin{minipage}[t]{0.48\columnwidth}
\begin{equation}
    \vec K=\tau_s^o\vec I-\rho_s\vec L,~~~
\end{equation}
\end{minipage}

\vspace{-12pt}

\noindent
\begin{minipage}[t]{0.48\columnwidth}
\begin{equation}
    \vec L=\mathbb{R}(\phi_2)\vec N,~~~~~~
\end{equation}
\end{minipage}\hfill
\begin{minipage}[t]{0.48\columnwidth}
\begin{equation}
    \vec M=\mathbb{R}(\phi_2)\vec K,~~~~~~~~
\end{equation}
\end{minipage}

\vspace{-12pt}

\noindent
\begin{minipage}[t]{0.48\columnwidth}
\begin{equation}
    \vec N=-\rho\vec M+\tau\vec P,~~~
\end{equation}
\end{minipage}
\hfill
\begin{minipage}[t]{0.48\columnwidth}
\begin{equation}
    \vec O=\tau\vec M+\rho\vec P.~~~
\end{equation}
\end{minipage}

\vspace{8pt}
\end{subequations}

The AC fields in the cavity satisfy the following equations, under our convention for evaluating the fields at the dielectric/vacuum interfaces (cf. Sec. \ref{sec: moving dielectric}):
\begin{subequations}
\begin{alignat}{2}
    \vec b &= \rho\bigl(\vec a- 2kx_\text{in}(\Omega)\vec A_\perp\bigr) + \tau\vec d , \\
    \vec c &= \tau\vec a - \rho\bigl(\vec d + 2kx_\text{in}(\Omega)\vec D_\perp\bigr), \\
    \vec d &= \mathbb{R}(\phi_1)\bigl(\exp(i\Omega L_t/c)\vec f - kx\vec F_\perp\bigr), \\
    \vec e &= \mathbb{R}(\phi_1)\bigl(\exp(i\Omega L_t/c)\vec c - kx\vec C_\perp\bigr), \\
    \vec f &= -\rho_s\vec e + \tau_s^o\vec h, \\
    \vec g &= \tau_s^i\vec e + \rho_s\vec h, \\
    \vec h &= \mathbb{R}(\bar{k}L_s)\exp(i\Omega nL_s/c)\vec j, \\
    \vec i &= \mathbb{R}(\bar{k}L_s)\exp(i\Omega nL_s/c)\vec g, \\
    \vec j &= \rho_s\vec i + \tau_s^i\vec l, \\
    \vec k &= \tau_s^o\vec i - \rho_s\vec l, \\
    \vec l &= \mathbb{R}(\phi_2)\bigl(\exp(i\Omega(L_c - L_t - L_s)/c)\vec n + kx\vec N_\perp\bigr) \\
    \vec m &= \mathbb{R}(\phi_2)\bigl(\exp(i\Omega(L_c-L_t-L_s)/c)\vec k + kx\vec K_\perp\bigr) \\
    \vec n &= -\rho\bigl(\vec m - 2kx_\text{end}(\Omega)\vec M_\perp\bigr) + \tau\vec p, \\
    \vec o &= \tau\vec m + \rho\bigl(\vec p+2kx_\text{end}(\Omega)\vec{P}_\perp\bigr).
\end{alignat}
\end{subequations}

Here, $L_c$, $L_t$, and $L_s$ are the macroscopic cavity length, macroscopic distance of the trap site from the input mirror, and sensor thickness. Also, we define $\phi_1\equiv \Delta L_t/c+\phi_f$, $\bar{k}\equiv n(k+\Delta/c)$, and $\phi_2\equiv \Delta(L_c-L_t-L_s)/c-\phi_f-nkL_s+\phi_e$ (our definition enforces $2\pi N$, $N\in\mathbb{N}$ phase for a full round trip of a zero-detuned beam when $\phi_e=0$). Furthermore, $\rho$ and $\tau$ are the reflectivity and transmissivity of the cavity mirrors, while the sensor reflectivity and ingoing and outgoing transmissivities are $\rho_s=(n-1)/(n+1)$, $\tau_s^i=2/(n+1)$, $\tau_s^o=2n/(n+1)$ with $n$ being the dielectric index of refraction. Finally, $x$ (with $\Omega$ dependence suppressed) is the displacement of the dielectric sensor, while $x_\text{in}(\Omega)$ and $x_\text{end}(\Omega)$ are the displacement of the input and end cavity mirrors, respectively.

Lastly, the equation for the motion of the sensor in full is, as described in Eq. \eqref{eq: force coefficient sum},

\begin{multline}
    \chi_\text{mech}^{-1}(\Omega)x(\Omega)=\sum_{z\in Z_\text{trap}}\frac{\partial }{\partial z}[\delta F_\text{trap}(\Omega)]z(\Omega)
    \\
    +\sum_{z\in Z_\text{probe}}\frac{\partial }{\partial z}[\delta F_\text{probe}(\Omega)]z(\Omega)+\delta F_\text{ext}(\Omega),
\end{multline}

where $Z_\text{trap}=\{x_\text{end},$ $x_\text{in},$ $a_{1,\text{trap}},$ $a_{2,\text{trap}},$ $p_{1,\text{trap}},$ $p_{2,\text{trap}}\}$ and $Z_\text{probe}=\{x,$ $x_\text{end},$ $x_\text{in},$ $a_{1,\text{probe}},$ $a_{2,\text{probe}},$ $p_{1,\text{probe}},$ $p_{2,\text{probe}}\}$ are the sets of valid perturbation entry ports and the force coefficients are calculated using the methods outlined in Sec. \ref{sec: sideband field forces}.

\section{Illustrative effect of cooling on a simplified noise budget}

\begin{figure}[!t]
    \centering
    \includegraphics[width=0.95\linewidth]{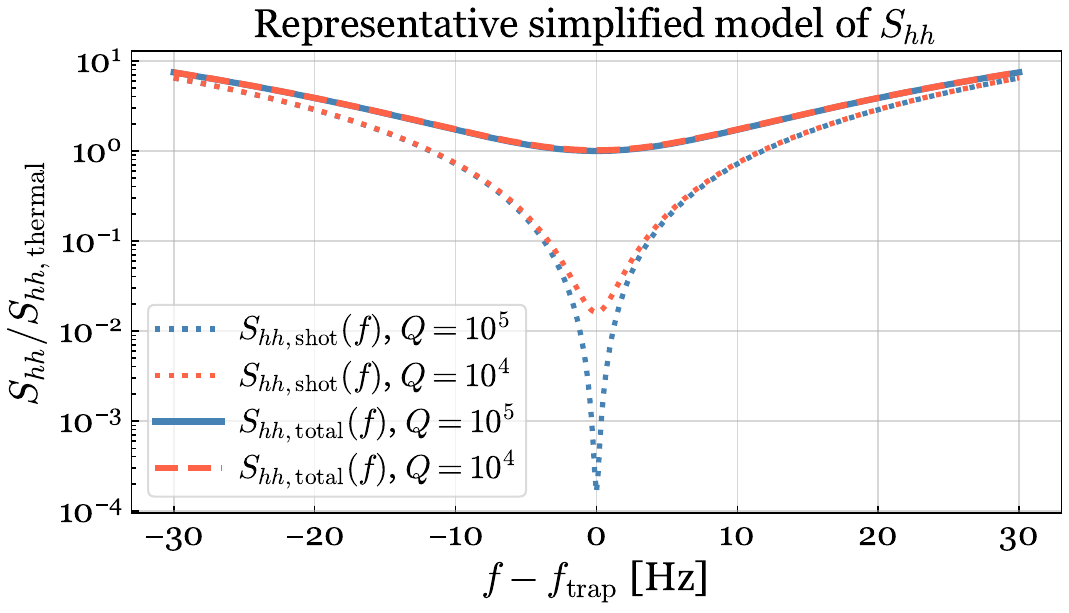}
    \caption{A representative model for the strain noise PSD, which consists of a constant contribution due to thermal force noise $S_{hh,\text{thermal}}(f)$ and a shot noise contribution $S_{hh,\text{shot}}(f)$ which behaves like the inverse square of $\chi_\text{mech}(f)$. Here, we have arbitrarily chosen parameters of $L=1$ m, $m=10^{-14}$ kg, $\Omega_\text{trap}=2\pi\times30$ kHz, $T_\text{eff}/Q_\text{eff}=3$ K, and $R=10^{-18}$. For two $Q$-factors $Q_\text{eff}=10^5$ and $10^4$, the shot noise contribution varies heavily in the vicinity of the trap frequency, but because the thermal contribution is dominant there, the overall noise PSD is essentially unaffected.}
    \label{fig: Shh model}
\end{figure}

\label{app: noise vs Q}
In this Appendix, we briefly discuss the effect of cooling the sensor on a rudimentary noise budget model. Consider a simplified system with just two noise sources: one which appears directly at the readout (originating from, say, quantum shot noise) and one which propagates to the readout via a force on the sensor from some thermal source. The measurement scheme for this detector relies on observing the motion of the sensor, so we can understand the role of cooling from the perspective of the spectral density of the noise in the sensor position.

We impose the well-motivated PSD for the thermal force noise $S_{FF,\text{thermal}}=k_B(T_\text{eff}/Q_\text{eff})m\Omega_\text{trap}$, where $T_\text{eff}/Q_\text{eff}$ is a constant. The resulting PSD for sensor displacement noise is simply $S_{xx,\text{thermal}}=S_{FF,\text{thermal}}|\chi_\text{mech}(\Omega)|^2$, where $\chi_\text{mech}$ is as in Eq. \ref{eq: chi mech} with $\Gamma\equiv\Omega_\text{trap}/Q_\text{eff}$. Meanwhile, the noise appearing directly in the readout can be converted to effective sensor displacement noise by dividing by the square of the sensor displacement to readout transfer function. The relevant frequency scales for this transfer function are purely optical in nature (rather than mechanical), so we can assume that it is essentially constant over the width of the narrow band of high sensitivity centered on the trap frequency, writing $S_{xx,\text{shot}}=R$.

Assuming the sensor is trapped very near the input mirror so that $L-x_0\approx L$, the total noise PSD in strain units is roughly
\begin{multline}
    S_{hh,\text{total}}(\Omega)\approx\frac{4}{mL^2\Omega_\text{trap}^4}\Bigr[k_B\frac{T_\text{eff}}{Q_\text{eff}}\Omega_\text{trap}\\
    +mR\Bigr(\frac{\Omega^2\Omega_\text{trap}^2}{Q_\text{eff}^2}+(\Omega^2-\Omega_\text{trap}^2)^2\Bigr)\Bigr].
\end{multline}
The minimum of $S_\text{hh,\text{total}}$ occurs at $\Omega=\Omega_\text{trap}$ with
\begin{equation}
    \text{min}(S_{hh,\text{total}})=\frac{4}{L^2}\Bigr(\frac{R}{Q_\text{eff}^2}+\frac{k_BT_\text{eff}/Q_\text{eff}}{m\Omega_\text{trap}^3}\Bigr).
\end{equation}
If $Q_\text{eff}$ is large so that the first term is subdominant, then the minimum becomes independent of the cooled $Q_\text{eff}$. Furthermore, when this condition (i.e., sufficiently large $Q_\text{eff}$) is met, the width of the narrow sensitivity band, defined by the separation between the frequencies which satisfy $S_{hh,\text{total}}(\Omega)=2\times\text{min}(S_{hh,\text{total}})$, is equal to $\sqrt{k_B(T_\text{eff}/Q_\text{eff})/mR\Omega_\text{trap}}$. When the system is cooled, $T_\text{eff}/Q_\text{eff}$ remains constant. 

Thus, under the reasonable assumption that the total strain noise is limited by the thermal force noise contribution at the resonant frequency \cite{LSD2012, LSD2020}, cooling does not impact either the minimum of the strain noise PSD or the width of the band of sensitivity --- see Fig. \ref{fig: Shh model}. As such, the cooling ability of the probe beam likely will not noticeably impact detector noise budget estimates.

\end{document}